\documentclass[12pt]{article}
\usepackage{amsmath,amssymb}
\usepackage{graphicx}
\newcommand{\eqdef}{\stackrel{\text{def}}{=}}
\newcommand{\eqdefrm}{\stackrel{\text{\rm def}}{=}}
\newcommand{\n}{\nonumber\\}
\newcommand{\bm}{\boldsymbol}

\newcommand{\ignore}[1]{}
\numberwithin{equation}{section}
\newcommand{\Romannumeral}[1]{\uppercase\expandafter{\romannumeral#1}}
\newcommand{\I}{\text{\Romannumeral{1}}}
\newcommand{\II}{\text{\Romannumeral{2}}}

\newcommand{\Irm}{\text{\rm\Romannumeral{1}}}
\newcommand{\IIrm}{\text{\rm\Romannumeral{2}}}

\newtheorem{lemma}{\bf Lemma}[section]
\newtheorem{prop}{\bf Proposition}[section]
\newtheorem{thm}{\bf Theorem}[section]

\newtheorem{conj}{\bf Conjecture}[section]
\newcounter{myremarkbangou}[section]
\renewcommand{\themyremarkbangou}{\arabic{section}.\arabic{myremarkbangou}}
\newcommand{\remark}{\refstepcounter{myremarkbangou}
\noindent{\bf Remark \themyremarkbangou\ }}
\newcommand{\heihoukon}{\raisebox{1mm}{$\sqrt{~~~}\,$}}

\newcommand{\cP}{\mathcal{P}}
\newcommand{\ccP}{\check{\mathcal{P}}}
\newcommand{\cM}{\mathcal{M}}
\newcommand{\cMt}{\widetilde{\mathcal{M}}}
\newcommand{\cN}{\mathcal{N}}

\newcommand{\llcN}{{\scriptscriptstyle\mathcal{N}}}
\newcommand{\cNt}{\widetilde{\mathcal{N}}}

\newcommand{\llcNt}{{\scriptscriptstyle\widetilde{\mathcal{N}}}}
\allowdisplaybreaks[4]

\begin{document}

\baselineskip=20pt

\newcommand{\preprint}{
\vspace*{-20mm}
   \begin{flushright}\normalsize \sf
    DPSU-21-2\\
  \end{flushright}}
\newcommand{\Title}[1]{{\baselineskip=26pt
  \begin{center} \Large \bf #1 \\ \ \\ \end{center}}}
\newcommand{\Author}{\begin{center}
  \large \bf Satoru Odake \end{center}}
\newcommand{\Address}{\begin{center}
     Faculty of Science, Shinshu University,
     Matsumoto 390-8621, Japan
   \end{center}}
\newcommand{\Accepted}[1]{\begin{center}
  {\large \sf #1}\\ \vspace{1mm}{\small \sf Accepted for Publication}
  \end{center}}

\preprint
\thispagestyle{empty}

\Title{Discrete Orthogonality Relations for the Multi-Indexed Orthogonal
Polynomials in Discrete Quantum Mechanics with Pure Imaginary Shifts}

\Author

\Address
\vspace{1cm}

\begin{abstract}
The discrete orthogonality relations for the multi-indexed orthogonal
polynomials in discrete quantum mechanics with pure imaginary shifts are
investigated. We show that the discrete orthogonality relations hold for the
case-(1) multi-indexed orthogonal polynomials of continuous Hahn, Wilson and
Askey-Wilson types, and conjecture their normalization constants.
\end{abstract}

\section{Introduction}
\label{sec:intro}

Ordinary orthogonal polynomials in one variable, $P_n(\eta)$
($n\in\mathbb{Z}_{\ge 0}$), are characterized by the three term recurrence
relations \cite{aar},
\begin{equation}
  \eta P_n(\eta)=A_nP_{n+1}(\eta)+B_nP_n(\eta)+C_nP_{n-1}(\eta)\quad
  (n\in\mathbb{Z}_{\ge 0}),
  \label{3trr}
\end{equation}
where $P_n(\eta)$ is a polynomial of degree $n$ in $\eta$ and $P_{-1}(\eta)=0$.
The hypergeometric orthogonal polynomials of the Askey scheme satisfy the second
order differential or difference equations \cite{kls}.
New types of orthogonal polynomials $P_{\mathcal{D},n}(\eta)$
($n\in\mathbb{Z}_{\ge 0}$), exceptional or multi-indexed orthogonal polynomials
\cite{gomez}--\cite{idQMcH}, satisfy the second order differential or
difference equations, but do not satisfy the three term recurrence relations
because of the missing degrees.
We distinguish the following two cases;
the set of missing degrees $\mathcal{I}=\mathbb{Z}_{\geq 0}\backslash
\{\text{deg}\,P_{\mathcal{D},n}(\eta)|n\in\mathbb{Z}_{\geq 0}\}$ is
case-(1): $\mathcal{I}=\{0,1,\ldots,\ell-1\}$, or
case-(2): $\mathcal{I}\neq\{0,1,\ldots,\ell-1\}$, where $\ell$ is a positive
integer. The situation of case-(1) is called stable in \cite{gkm11}.
Our study of orthogonal polynomials is based on the quantum mechanical
formulation: ordinary quantum mechanics (oQM), discrete quantum mechanics
with pure imaginary shifts (idQM) \cite{os13}--\cite{os24} and discrete
quantum mechanics with real shifts (rdQM) \cite{os12}--\cite{os34}.
The Schr\"odinger equation of oQM is a differential equation and that of dQM is
a difference equation.
We deform exactly solvable quantum mechanical systems by
multi-step Darboux transformations and obtain multi-indexed polynomials as
eigenfunctions of the deformed systems.
They are polynomials in the sinusoidal coordinate $\eta(x)$ \cite{os7,os14},
$P_{\mathcal{D},n}(\eta(x))$, where $x$ is the coordinate of the quantum system.
The case-(1) multi-indexed polynomials are obtained by taking the virtual state
wavefunctions as seed solutions of the Darboux transformations.
When the eigenfunctions are taken as seed solutions \cite{krein,adler},
the resulting multi-indexed polynomials are case-(2), and we call them
Krein-Adler type multi-indexed polynomials.

Any ordinary orthogonal polynomials $P_n(\eta)$ satisfy the discrete orthogonal
relations.
Let us fix a positive integer $\cN$ and denote the zeros of $P_{\cN}(\eta)$ as
$\eta_j$ ($j=1,2,\ldots,\cN$). Then the following discrete orthogonal relations
hold \cite{aar}:
\begin{equation}
  \sum_{j=1}^{\cN}\text{sgn}(C_{\cN})\frac{P'_{\cN}(\eta_j)}{P_{\cN-1}(\eta_j)}
  \cdot\frac{P_n(\eta_j)}{P'_{\cN}(\eta_j)}\frac{P_m(\eta_j)}{P'_{\cN}(\eta_j)}
  =|C_{\cN}|\frac{h_n}{h_{\cN}}\delta_{nm}\quad(0\leq n,m<\cN),
  \label{doorg}
\end{equation}
where $f'(x)=\frac{df(x)}{dx}$, and $\text{sgn}(C_{\cN})$ is inserted for the
positivity of the weight factor.
Since the multi-indexed orthogonal polynomials $P_{\mathcal{D},n}(\eta)$ are
deformations of the ordinary orthogonal polynomials $P_n(\eta)$, it is expected
that the multi-indexed polynomials also satisfy the discrete orthogonal
relations. We naively expected the following:
\begin{equation}
  \sum_{j=1}^{\cNt}\frac{P'_{\mathcal{D},\llcN}(\eta_j)}
  {P_{\mathcal{D},\llcN-1}(\eta_j)}
  \cdot\frac{P_{\mathcal{D},n}(\eta_j)}{P'_{\mathcal{D},\llcN}(\eta_j)}
  \frac{P_{\mathcal{D},m}(\eta_j)}{P'_{\mathcal{D},\llcN}(\eta_j)}
  \stackrel{?}{=}0\quad(n\neq m,\ 0\leq n,m<\cN),
  \label{donaive}
\end{equation}
where $\cNt=\deg P_{\mathcal{D},\llcN}(\eta)$.
At first, we approached this problem by numerical calculation for the case-(1)
multi-indexed polynomials of Laguerre, Jacobi, Wilson and Askey-Wilson types.
It suggested that the discrete orthogonal relations hold for the multi-indexed
polynomials, but \eqref{donaive} does not hold, namely the weight is not given
by $P'_{\mathcal{D},\llcN}(\eta_j)/P_{\mathcal{D},\llcN-1}(\eta_j)$.
In a private communication (2017), Sasaki informed the author that the
weights for the multi-indexed Laguerre (L) and Jacobi (J) (and Hermite (H))
polynomials are given by $1/c_2(\eta_j)$, based on the perturbations around
the zeros of orthogonal polynomials \cite{s14}.
Here $c_2(\eta)$ is given by (2.132) of \cite{os24}
\begin{equation}
  c_2(\eta)=\left\{
  \begin{array}{ll}
  \frac14&:\text{H}\\
  \eta&:\text{L}\\
  1-\eta^2&:\text{J}
  \end{array}\right..
  \label{c2eta}
\end{equation}
In fact,
$P_n(\eta)=H_n(\eta),L^{(g-\frac12)}_n(\eta),P^{(g-\frac12,h-\frac12)}_n(\eta)$
for H, L, J cases satisfy
\begin{align*}
  \text{H}:\ \ &P'_n(\eta)-2nP_{n-1}(\eta)=0,\\
  \text{L}:\ \ &\eta P'_n(\eta)+(n+g-\tfrac12)P_{n-1}(\eta)=nP_n(\eta),\\
  \text{J}:\ \ &(2n+g+h-1)(1-\eta^2)P'_n(\eta)
  -2(n+g-\tfrac12)(n+h-\tfrac12)P_{n-1}(\eta)\n
  &=-n\bigl(h-g+(2n+g+h-1)\eta\bigr)P_n(\eta),
\end{align*}
and these imply
\begin{equation*}
  \frac{P'_{\cN}(\eta_j)}{P_{\cN-1}(\eta_j)}=\left\{
  \begin{array}{ll}
  \frac14\times 8\cN&:\text{H}\\[4pt]
  {\displaystyle\frac{1}{\eta_j}\times(-1)(\cN+g-\tfrac12)}&:\text{L}\\[4pt]
  {\displaystyle\frac{1}{1-\eta^2_j}\times
  \frac{2(\cN+g-\frac12)(\cN+h-\frac12)}{2\cN+g+h-1}}&:\text{J}
  \end{array}\right..
\end{equation*}
We verified this weight $1/c_2(\eta_j)$ gives the discrete orthogonality
relations for the case-(1) multi-indexed Laguerre and Jacobi polynomials by
numerical calculation.
For the case-(1) multi-indexed Wilson and Askey-Wilson polynomials, however,
we could not find analytical expression of the weight at that time.

Recently Ho and Sasaki showed that the discrete orthogonality relations with
the weight $1/c_2(\eta_j)$ hold for the multi-indexed orthogonal polynomials in
oQM: the case-(1) multi-indexed Laguerre and Jacobi polynomials, and the
Krein-Adler type multi-indexed polynomials based on the Hermite, Laguerre and
Jacobi polynomials \cite{hs19}.
Motivated by the perturbations around the zeros of orthogonal polynomials
\cite{s14}, they consider a matrix $\cMt$ and its symmetric version $\cM$.
Orthogonality of the eigenvectors of this symmetric matrix $\cM$ implies
the discrete orthogonality relations for the multi-indexed polynomials.

In this paper, we consider the discrete orthogonality relations for the
multi-indexed orthogonal polynomials in idQM.
The strategy is the same as Ho-Sasaki. By considering a matrix $\cMt$ and its
symmetric version $\cM$, we show that the discrete orthogonality relations hold
for the case-(1) multi-indexed polynomials of continuous Hahn, Wilson and
Askey-Wilson types.

This paper is organized as follows.
In section \ref{sec:do} we present a general theory of the discrete
orthogonality relations for orthogonal polynomials satisfying differential or
difference equations.
In section \ref{sec:miop} the discrete orthogonality relations for the case-(1)
multi-indexed orthogonal polynomials of continuous Hahn, Wilson and
Askey-Wilson types are presented.
Section \ref{sec:summary} is for a summary and comments.
In Appendix \ref{app:pr:id} Proposition\,\ref{prop:id} is proved.

\section{Discrete Orthogonality Relations}
\label{sec:do}

In this section we present a general theory of the discrete orthogonality
relations for orthogonal polynomials satisfying differential or difference
equations.
The basic idea is given in \cite{hs19}.

Let us consider $\check{P}_n(x)=P_n(\eta(x))$
($n\in\mathbb{Z}_{\geq 0}$), where $x$ is a coordinate of some quantum system
(physical range: $x_1\leq x\leq x_2$), $\eta(x)$ is the sinusoidal coordinate
\cite{os7,os14}, and $P_n(\eta)$ is a polynomial in $\eta$.
We assume that $P_n(y)$'s are orthogonal polynomials,
\begin{equation}
  \int_{x_1}^{x_2}\!dx\,\psi(x)^2\check{P}_n(x)\check{P}_m(x)
  =h_n\delta_{nm}\quad(n,m\in\mathbb{Z}_{\geq 0}),
  \label{intcPncPm}
\end{equation}
and satisfy a differential or difference equation,
\begin{equation}
  \widetilde{\mathcal{H}}\check{P}_n(x)=\mathcal{E}_n\check{P}_n(x)
  \quad(n\in\mathbb{Z}_{\geq 0}).
  \label{HtcPn}
\end{equation}
Here $\psi(x)^2$ is a weight function, $h_n$'s are normalization constants
($h_n>0$), a differential or difference operator $\widetilde{\mathcal{H}}$ is
a transformed Hamiltonian (`true' Hamiltonian is
$\mathcal{H}=\psi(x)\circ\widetilde{\mathcal{H}}\circ\psi(x)^{-1}$),
and $\mathcal{E}_n$'s are its energy eigenvalues.
We assume $\deg P_n<\deg P_m$ and $\mathcal{E}_n<\mathcal{E}_m$ for $n<m$.
The degree of $P_n$ is $\deg P_n=n$ for ordinary orthogonal polynomials, but
$\deg P_n\geq n$ for multi-indexed orthogonal polynomials.

Let us fix a non-negative integer $\cN$ and set $\cNt=\deg P_{\cN}$.
We denote the zeros of $P_{\cN}(\eta)$ as $\eta^{[\llcN]}_j$
($j=1,2,\ldots,\cNt$), which may be complex, and assume that they are simple.
The sinusoidal coordinates considered in this paper are $\eta(x)=x,x^2$ and
$\cos x$ (or $\cos 2x$).
The sinusoidal coordinate $\eta(x)$ and the coordinate $x$ have a one-to-one
correspondence for the physical value of $x$, but this may not be the case
for unphysical value of $x$.
We fix $x^{[\llcN]}_j$ uniquely, which gives
$\eta^{[\llcN]}_j=\eta(x^{[\llcN]}_j)$, by requiring
$x_1\leq\text{Re}\,x^{[\llcN]}_j\leq x_2$.

Let us assume the existence of $\ccP_a(x)=\cP_a(\eta(x))$
($a=1,2,\ldots,\cNt$) that satisfy the following conditions:
\begin{equation}
  \begin{array}{ll}
  \text{(\romannumeral1)}
  &\!\!:\ \text{$\cP_a(\eta)$ : a polynomial in $\eta$},\\[2pt]
  \text{(\romannumeral2)}&\!\!:\ \deg\cP_a<\cNt,\\[2pt]
  \text{(\romannumeral3)}
  &\!\!:\ \widetilde{\mathcal{H}}\ccP_a(x)\bigl|_{x=x^{[\llcN]}_j}\,=
  \mathcal{E}^{\cP}_a\cP_a(\eta^{[\llcN]}_j),\\[6pt]
  \text{(\romannumeral4)}
  &\!\!:\ \mathcal{E}^{\cP}_a\neq\mathcal{E}^{\cP}_b\ \ (a\neq b).
  \end{array}
  \label{cPacond}
\end{equation}
The condition $\deg\cP_a<\deg\cP_b$ for $a<b$ is not imposed.
Note that $P_n(\eta)$ ($n=0,1,\ldots,\cN-1$) satisfy the conditions
\eqref{cPacond} with $\mathcal{E}^{\cP}_a=\mathcal{E}_n$.
Since $\cP_a$ is a polynomial of $\deg\cP_a<\cNt$,
it is expressed as
\begin{equation}
  \cP_a(\eta)=\sum_{j=1}^{\cNt}
  \frac{\prod\limits_{\genfrac{}{}{0pt}{}{l=1}{l\neq j}}^{\llcNt}
  (\eta-\eta^{[\llcN]}_l)}
  {\prod\limits_{\genfrac{}{}{0pt}{}{l=1}{l\neq j}}^{\llcNt}
  (\eta^{[\llcN]}_j-\eta^{[\llcN]}_l)}\,
  \cP_a(\eta^{[\llcN]}_j),
  \label{cPaexp}
\end{equation}
because the both sides agree at $\cNt$ points $\eta=\eta^{[\llcN]}_j$.
Note that this can be rewritten as
\begin{equation}
  \cP_a(\eta)=\sum_{j=1}^{\cNt}
  \frac{c^P_{\cN}\prod\limits_{l=1}^{\llcNt}
  (\eta-\eta^{[\llcN]}_l)}
  {(\eta-\eta^{[\llcN]}_j)c^P_{\cN}
  \prod\limits_{\genfrac{}{}{0pt}{}{l=1}{l\neq j}}^{\llcNt}
  (\eta^{[\llcN]}_j-\eta^{[\llcN]}_l)}\,
  \cP_a(\eta^{[\llcN]}_j)
  =\sum_{j=1}^{\cNt}
  \frac{P_{\cN}(\eta)}{\eta-\eta^{[\llcN]}_j}
  \frac{\cP_a(\eta^{[\llcN]}_j)}{P'_{\cN}(\eta^{[\llcN]}_j)},
  \label{cPaexp2}
\end{equation}
where $P_{\cN}(\eta)=c^P_{\cN}\eta^{\llcNt}+(\text{lower degree terms})$ and
$P'_{\cN}(\eta)=\frac{d}{d\eta}P_{\cN}(\eta)$.
By replacing $j$ with $k$ in \eqref{cPaexp},
the action of $\widetilde{\mathcal{H}}$ on $\check{P}_a(x)$ is
\begin{equation}
  \widetilde{\mathcal{H}}\ccP_a(x)
  =\sum_{k=1}^{\cNt}\widetilde{\mathcal{H}}
  \prod\limits_{\genfrac{}{}{0pt}{}{l=1}{l\neq k}}^{\llcNt}
  \bigl(\eta(x)-\eta^{[\llcN]}_l\bigr)
  \times\frac{\cP_a(\eta^{[\llcN]}_k)}
  {\prod\limits_{\genfrac{}{}{0pt}{}{l=1}{l\neq k}}^{\llcNt}
  (\eta^{[\llcN]}_k-\eta^{[\llcN]}_l)}.
  \label{HtccPa}
\end{equation}
Let us evaluate this equation at $x=x^{[\llcN]}_j$.
By the condition (\romannumeral3) in \eqref{cPacond}, we obtain
\begin{equation}
  \mathcal{E}^{\cP}_a\cP_a(\eta^{[\llcN]}_j)
  =\sum_{k=1}^{\cNt}\cMt_{jk}
  \prod\limits_{\genfrac{}{}{0pt}{}{l=1}{l\neq j}}^{\llcNt}
  (\eta^{[\llcN]}_j-\eta^{[\llcN]}_l)
  \times\frac{\cP_a(\eta^{[\llcN]}_k)}
  {\prod\limits_{\genfrac{}{}{0pt}{}{l=1}{l\neq k}}^{\llcNt}
  (\eta^{[\llcN]}_k-\eta^{[\llcN]}_l)},
  \label{EPcPa=}
\end{equation}
where $\cMt_{jk}$ is defined by
\begin{equation}
  \cMt_{jk}\eqdef\frac{\widetilde{\mathcal{H}}
  \prod\limits_{\genfrac{}{}{0pt}{}{l=1}{l\neq k}}^{\llcNt}
  \bigl(\eta(x)-\eta^{[\llcN]}_l\bigr)\biggl|_{x=x^{[\llcN]}_j}}
  {\prod\limits_{\genfrac{}{}{0pt}{}{l=1}{l\neq j}}^{\llcNt}
  (\eta^{[\llcN]}_j-\eta^{[\llcN]}_l)}
  \quad(j,k=1,2,\ldots,\cNt).
  \label{cMtdef}
\end{equation}
By dividing \eqref{EPcPa=} by
$\prod\limits_{\genfrac{}{}{0pt}{}{l=1}{l\neq j}}^{\llcNt}
(\eta^{[\llcN]}_j-\eta^{[\llcN]}_l)$, it becomes
\begin{equation}
  \cMt\tilde{v}^{(a)}=\mathcal{E}^{\cP}_a\tilde{v}^{(a)},
  \label{cMtvta=}
\end{equation}
where a $\cNt\times\cNt$ matrix $\cMt$ and $\cNt$-dimensional column vectors
$\tilde{v}^{(a)}$ are defined by
\begin{equation}
  \cMt=(\cMt_{jk}),\ \ \tilde{v}^{(a)}=(\tilde{v}^{(a)}_j),
  \ \ \tilde{v}^{(a)}_j\eqdef\frac{\cP_a(\eta^{[\llcN]}_j)}
  {\prod\limits_{\genfrac{}{}{0pt}{}{l=1}{l\neq j}}^{\llcNt}
  (\eta^{[\llcN]}_j-\eta^{[\llcN]}_l)}
  =c^P_{\cN}\frac{\cP_a(\eta^{[\llcN]}_j)}{P'_{\cN}(\eta^{[\llcN]}_j)}.
\end{equation}
By a similarity transformation in terms of a non-singular diagonal matrix
$G=\text{diag}(g_1,\ldots,g_{\llcNt})$, we define a matrix $\cM$ and
vectors $v^{(a)}$ as
\begin{equation}
  \cM\eqdef G^{-1}\cMt G,\ v^{(a)}\eqdef G^{-1}\tilde{v}^{(a)}
  \ \Bigl(\Rightarrow\cM_{jk}=\frac{g_k}{g_j}\cMt_{jk},
  \ v^{(a)}_j=\frac{1}{g_j}\tilde{v}^{(a)}_j,
  \ \cM v^{(a)}=\mathcal{E}^{\cP}_av^{(a)}\Bigr).\!\!\!
  \label{cM}
\end{equation}
We assume the existence of $G$ such that $\cM$ is symmetric,
\begin{equation}
  \cM_{jk}=\cM_{kj}.
  \label{cMsym}
\end{equation}

Then we have the following theorem.
\begin{thm}\label{thm:do}
For $\check{P}_n(x)$ satisfying \eqref{HtcPn},
$\cP_a(\eta)$ satisfying \eqref{cPacond} and $G$ giving \eqref{cMsym},
we have the discrete orthogonality relations:
\begin{equation}
  \sum_{j=1}^{\cNt}\frac{1}{g_j^2}
  \frac{\cP_a(\eta^{[\llcN]}_j)}{P'_{\cN}(\eta^{[\llcN]}_j)}
  \frac{\cP_b(\eta^{[\llcN]}_j)}{P'_{\cN}(\eta^{[\llcN]}_j)}
  =k^{[\llcN]}_a\delta_{ab}\quad(a,b=1,2,\ldots,\cNt),
  \label{do}
\end{equation}
where $k^{[\llcN]}_a$'s are normalization constants.
\end{thm}
Proof: Let us consider a vector space $\mathbb{C}^{\llcNt}\ni v=(v_j)$ and
a bilinear form $\langle\ ,\ \rangle:
\mathbb{C}^{\llcNt}\times\mathbb{C}^{\llcNt}\to\mathbb{C}$,
\begin{equation*}
  \langle v,w\rangle\eqdef\sum_{j=1}^{\cNt}v_jw_j\quad
  (v,w\in\mathbb{C}^{\llcNt}).
\end{equation*}
Since $\cM$ is symmetric, we have
$\langle\cM v,w\rangle=\langle v,\cM w\rangle$.
For $v^{(a)}$ and $v^{(b)}$, we have
\begin{equation*}
  \langle v^{(a)},v^{(b)}\rangle
  =\sum_{j=1}^{\cNt}\frac{1}{g_j^2}\tilde{v}^{(a)}_j\tilde{v}^{(b)}_j
  =(c^P_{\cN})^2\sum_{j=1}^{\cNt}\frac{1}{g_j^2}
  \frac{\cP_a(\eta^{[\llcN]}_j)}{P'_{\cN}(\eta^{[\llcN]}_j)}
  \frac{\cP_b(\eta^{[\llcN]}_j)}{P'_{\cN}(\eta^{[\llcN]}_j)},
\end{equation*}
and
\begin{equation*}
  \mathcal{E}^{\cP}_a\langle v^{(a)},v^{(b)}\rangle
  =\langle\cM v^{(a)},v^{(b)}\rangle
  =\langle v^{(a)},\cM v^{(b)}\rangle
  =\mathcal{E}^{\cP}_b\langle v^{(a)},v^{(b)}\rangle,
\end{equation*}
which implies $\langle v^{(a)},v^{(b)}\rangle=0$ for $a\neq b$.
By setting $k^{[\llcN]}_a$ as
$k^{[\llcN]}_a=(c^P_{\cN})^{-2}\langle v^{(a)},v^{(a)}\rangle$, we obtain
\eqref{do}.
\hfill\fbox{}

\medskip
\remark\label{rem:posi}
The weight $1/g_j^2$ in \eqref{do} may not be positive and may be
complex, and the normalization constant $k^{[\llcN]}_a$ may not be positive.

\remark\label{rem:kNa}
The proof uses \eqref{HtcPn} but not \eqref{intcPncPm} explicitly,
and the theorem states nothing about the properties of $k^{[\llcN]}_a$.
The theorem states nothing for $\cNt=0,1$ cases either.
Let us consider $\cP_a=P_n$ case.
By using \eqref{intcPncPm} and \eqref{cPaexp2}, we have
\begin{align}
  &\quad\int_{x_1}^{x_2}\!dx\,\psi(x)^2\check{P}_n(x)\check{P}_m(x)
  =h_n\delta_{nm}\n
  &=\sum_{j=1}^{\cNt}\sum_{k=1}^{\cNt}
  \frac{P_n(\eta^{[\llcN]}_j)}{P'_{\cN}(\eta^{[\llcN]}_j)}
  \frac{P_m(\eta^{[\llcN]}_k)}{P'_{\cN}(\eta^{[\llcN]}_k)}
  \int_{x_1}^{x_2}\!dx\,\psi(x)^2
  \frac{\check{P}_{\cN}(x)}{\eta(x)-\eta^{[\llcN]}_j}
  \frac{\check{P}_{\cN}(x)}{\eta(x)-\eta^{[\llcN]}_k}.
\end{align}
Since we know Theorem\,\ref{thm:do}, we naively expect the following equation,
\begin{equation}
  \int_{x_1}^{x_2}\!dx\,\psi(x)^2
  \frac{\check{P}_{\cN}(x)}{\eta(x)-\eta^{[\llcN]}_j}
  \frac{\check{P}_{\cN}(x)}{\eta(x)-\eta^{[\llcN]}_k}
  \stackrel{?}{=}0\quad(j\neq k).
\end{equation}
However, numerical calculation shows that this equation holds for ordinary
orthogonal polynomials, but does not for the multi-indexed orthogonal
polynomials.

\remark\label{rem:hs}
The Ho-Sasaki's matrices in \cite{hs19}, $\cMt^{\text{HS}}$ and
$\cM^{\text{HS}}$, correspond to $\cMt-\mathcal{E}_{\cN}\bm{1}$ and
$-\cM+\mathcal{E}_{\cN}\bm{1}$, respectively.
Their matrices are motivated by the perturbation around the zeros of orthogonal
polynomials \cite{s14} and the scalar matrix $\mathcal{E}_{\cN}\bm{1}$ is
subtracted.

\remark\label{rem:c81}
During the peer review process, we learned about the paper \cite{c81} from
the referee's comments.
The matrix $Z=(Z_{jk})_{j,k=1,2,\ldots,n}$
$Z_{jk}=\delta_{jk}\sum\limits_{\substack{l=1\\l\neq j}}^n(x_j-x_l)^{-1}
+(1-\delta_{jk})(x_j-x_k)^{-1}$
($x_j$ : arbitrary), and related topics are investigated \cite{c81}.
The results of \cite{c81} are interesting and important.
The logic of Theorem \ref{thm:do} is the same as that of \cite{hs19}, and both
follow the flow of \cite{c81}, with a partial extension with respect to the
missing degrees.
In the application to orthogonal polynomials given in \S\,2 of \cite{c81},
the three term recurrence relations and the properties obtained from them are
used.
As mentioned in \S\,\ref{sec:intro}, the multi-indexed orthogonal polynomials
considered in \cite{hs19} and this paper do not satisfy the three term
recurrence relations.
So the results given in \S\,2 of \cite{c81} can not be applied, and
the differential equations (for \cite{hs19}) and difference equations (for
this paper) play an important role.
This matrix $Z$ can be considered as a differential operator $\frac{d}{dx}$
in some situation (Remark : $x$ in \cite{c81} corresponds to $\eta$ in this
paper).
By applying this to some polynomial systems satisfying differential equations,
various interesting results are presented \cite{c81}.
The multi-indexed orthogonal polynomials in idQM satisfy the difference
equations but not the differential equations.
So the results of \cite{c81} can not be applied to them.
It is an interesting problem to generalize the result of \cite{c81} and
study certain matrix which may be interpreted as a difference operator
for $x$ in some situation.

\medskip

We present examples of Theorem\,\ref{thm:do} in the following sections.

\section{Multi-Indexed Orthogonal Polynomials}
\label{sec:miop}

In this section, after recapitulating the case-(1) multi-indexed orthogonal
polynomials in discrete quantum mechanics with pure imaginary shifts,
we show their discrete orthogonality relations and conjecture the normalization
constants.

The notations $P_n$ and $\widetilde{\mathcal{H}}$ in \S\,\ref{sec:do}
correspond to $P_{\mathcal{D},n}$ and $\widetilde{\mathcal{H}}_{\mathcal{D}}$
in this section, respectively.

\subsection{Discrete quantum mechanics with pure imaginary shifts}
\label{sec:idQM}

Let us recapitulate the discrete quantum mechanics with pure imaginary shifts
(idQM) \cite{os13,os24}.
The dynamical variables of idQM are the real coordinate $x$
($x_1\leq x\leq x_2$)
and the conjugate momentum $p=-i\frac{d}{dx}$, which are governed by the
following factorized positive semi-definite Hamiltonian:
\begin{align}
  &\mathcal{H}\eqdef\sqrt{V(x)}\,e^{\gamma p}\sqrt{V^*(x)}
  +\!\sqrt{V^*(x)}\,e^{-\gamma p}\sqrt{V(x)}
  -V(x)-V^*(x)=\mathcal{A}^{\dagger}\mathcal{A},
  \label{H}\\
  &\mathcal{A}\eqdef i\bigl(e^{\frac{\gamma}{2}p}\sqrt{V^*(x)}
  -e^{-\frac{\gamma}{2}p}\sqrt{V(x)}\,\bigr),\quad
  \mathcal{A}^{\dagger}\eqdef-i\bigl(\sqrt{V(x)}\,e^{\frac{\gamma}{2}p}
  -\sqrt{V^*(x)}\,e^{-\frac{\gamma}{2}p}\bigr).
\end{align}
Here the potential function $V(x)$ is an analytic function of $x$ and
$\gamma$ is a real constant.
The $*$-operation on an analytic function $f(x)=\sum_na_nx^n$
($a_n\in\mathbb{C}$) is defined by $f^*(x)=\sum_na_n^*x^n$, in which
$a_n^*$ is the complex conjugation of $a_n$.
Note that $\heihoukon$ is a square root as a complex function.
The Schr\"{o}dinger equation
\begin{equation}
  \mathcal{H}\phi_n(x)=\mathcal{E}_n\phi_n(x)
  \quad(n\in\mathbb{Z}_{\geq 0}),
  \label{Hphin=}
\end{equation}
is an analytic difference equation with pure imaginary shifts.
The inner product of two functions $f(x)$ and $g(x)$ is given by
$(f,g)\eqdef\int_{x_1}^{x_2}\!dx\,f^*(x)g(x)$.
The hermiticity of $\mathcal{H}$, $(f,\mathcal{H}g)=(\mathcal{H}f,g)$,
depends on singularities of some functions in the rectangular domain
$D_{\gamma}$ \cite{os13,os14,os27},
\begin{equation}
  D_{\gamma}\eqdef\bigl\{x\in\mathbb{C}\bigm|x_1\leq\text{Re}\,x\leq x_2,
  |\text{Im}\,x|\leq\tfrac12|\gamma|\bigr\}.
  \label{Dgamma}
\end{equation}
The eigenfunctions $\phi_n(x)$ can be chosen `real', $\phi_n^*(x)=\phi_n(x)$,
and the orthogonality relations read
\begin{equation}
  (\phi_n,\phi_m)=h_n\delta_{nm}\ \ (n,m\in\mathbb{Z}_{\geq 0}),
  \quad 0<h_n<\infty.
  \label{ortho}
\end{equation}

We consider the idQM systems whose eigenfunctions $\phi_n(x)$ \eqref{Hphin=}
have the following form:
\begin{equation}
  \phi_n(x)=\phi_0(x)\check{P}_n(x),\quad
  \check{P}_n(x)\eqdef P_n\bigl(\eta(x)\bigr)\quad
  (n\in\mathbb{Z}_{\geq 0}),
  \label{phin=}
\end{equation}
where $\eta(x)$ is a sinusoidal coordinate \cite{os7,os14} and
$P_n(\eta)$ is an orthogonal polynomial of degree $n$ in $\eta$ and satisfies
$\check{P}^*_n(x)=\check{P}_n(x)$.
The energy eigenvalues satisfy
$0=\mathcal{E}_0 <\mathcal{E}_1 < \mathcal{E}_2 < \cdots$.
As a polynomial $P_n(\eta)$, we consider the continuous Hahn (cH), Wilson (W)
and Askey-Wilson (AW) polynomials etc., which are members of the Askey-scheme
of hypergeometric orthogonal polynomials \cite{kls}.
We call the idQM system by the name of the orthogonal polynomial:
continuous Hahn system, Wilson system, Askey-Wilson system etc.
The similarity transformation in terms of the ground state wavefunction gives
the difference operator $\widetilde{\mathcal{H}}$, which acts on the polynomial
eigenfunctions and is square root free,
\begin{align}
  &\widetilde{\mathcal{H}}\eqdef
  \phi_0(x)^{-1}\circ\mathcal{H}\circ\phi_0(x)
  =V(x)(e^{\gamma p}-1) +V^*(x)(e^{-\gamma p}-1),
  \label{Ht}\\
  &\widetilde{\mathcal{H}}\check{P}_n(x)=\mathcal{E}_n\check{P}_n(x)
  \quad(n\in\mathbb{Z}_{\geq 0}).
  \label{HtP=EP}
\end{align}

Concrete idQM systems have a set of parameters
$\bm{\lambda}=(\lambda_1,\lambda_2,\ldots)$ and various quantities depend on
them. If necessary, their dependence is expressed like,
$f=f(\bm{\lambda})$, $f(x)=f(x;\bm{\lambda})$.
For cH, W and AW systems, the parameters are
$\bm{\lambda}=(\lambda_1,\lambda_2,\lambda_3,\lambda_4)$
($\lambda_i\in\mathbb{C}$, $\text{Re}\,\lambda_i>0$) and satisfy
\begin{align}
  \text{cH}\ &:\ \lambda_3=\lambda_1^*,\ \ \lambda_4=\lambda_2^*,
  \label{condcH}\\
  \text{W,\ AW}\ &:\ \{\lambda_1^*,\lambda_2^*,\lambda_3^*,\lambda_4^*\}
  =\{\lambda_1,\lambda_2,\lambda_3,\lambda_4\}\ \ (\text{as a set}),
  \label{condWAW}
\end{align}
We remark that $\bm{\lambda}$ for cH is taken as
$\bm{\lambda}=(\lambda_1,\lambda_2)$ in \cite{idQMcH} because of \eqref{condcH}.
The AW system contains the parameter $q$ ($0<q<1$).

The data for $V(x)$, $\mathcal{E}_n$, $\phi_0(x)$, $P_n(\eta)$, $\eta(x)$,
$h_n$, $\gamma$
are given in \S\,2.2 of \cite{idQMcH} and (2.25)--(2.26) of \cite{rrmiop}.
The parameters are $\bm{\lambda}=(a_1,a_2,a_3,a_4)$ for cH and W
($b_1=a_1+a_2+a_3+a_4$), and $q^{\bm{\lambda}}=(a_1,a_2,a_3,a_4)$ for AW
($b_4=a_1a_2a_3a_4$).

\subsection{Darboux transformations}
\label{sec:DT}

The exactly solvable idQM systems in \S\,\ref{sec:idQM} can be deformed by
the multi-step Darboux transformations.
We consider the Darboux transformations with virtual state wavefunctions as
seed solutions.
The $M$-step Darboux transformations are as follows.
Take $M$ virtual state wave functions,
$\tilde{\phi}_{d_1}(x),\tilde{\phi}_{d_2}(x),\ldots,\tilde{\phi}_{d_M}(x)$
($d_j$ : mutually distinct),
which are solutions of the Schr\"{o}dinger equation,
\begin{equation}
  \mathcal{H}\tilde{\phi}_{\text{v}}(x)
  =\tilde{\mathcal{E}}_{\text{v}}\tilde{\phi}_{\text{v}}(x),\quad
  \tilde{\mathcal{E}}_{\text{v}}<0.
  \label{vs}
\end{equation}
Note that $\tilde{\phi}^*_{\text{v}}(x)=\tilde{\phi}_{\text{v}}(x)$.
In the $s$-step ($1\leq s\leq M$), we have \cite{os27}
($s=0$ : $\mathcal{H}=\hat{\mathcal{A}}_{d_1}^{\dagger}
\hat{\mathcal{A}}_{d_1}+\tilde{\mathcal{E}}_{d_1}$)
\begin{align}
  &\mathcal{H}_{d_1\ldots d_s}\eqdef
  \hat{\mathcal{A}}_{d_1\ldots d_s}\hat{\mathcal{A}}_{d_1\ldots d_s}^{\dagger}
  +\tilde{\mathcal{E}}_{d_s}
  =\hat{\mathcal{A}}_{d_1\ldots d_sd_{s+1}}^{\dagger}
  \hat{\mathcal{A}}_{d_1\ldots d_sd_{s+1}}+\tilde{\mathcal{E}}_{d_{s+1}},
  \label{Hd1..ds}\\
  &\hat{\mathcal{A}}_{d_1\ldots d_s}\eqdef
  i\bigl(e^{\frac{\gamma}{2}p}\sqrt{\hat{V}_{d_1\ldots d_s}^*(x)}
  -e^{-\frac{\gamma}{2}p}\sqrt{\hat{V}_{d_1\ldots d_s}(x)}\,\bigr),\n
  &\hat{\mathcal{A}}_{d_1\ldots d_s}^{\dagger}\eqdef
  -i\bigl(\sqrt{\hat{V}_{d_1\ldots d_s}(x)}\,e^{\frac{\gamma}{2}p}
  -\sqrt{\hat{V}_{d_1\ldots d_s}^*(x)}\,e^{-\frac{\gamma}{2}p}\bigr),
  \label{Ahd1..ds}\\
  &\hat{V}_{d_1\ldots d_s}(x)\eqdef
  \sqrt{V(x-i\tfrac{s-1}{2}\gamma)V^*(x-i\tfrac{s+1}{2}\gamma)}\n
  &\phantom{\hat{V}_{d_1\ldots d_s}(x)\eqdef}\times
  \frac{\text{W}_{\gamma}[\tilde{\phi}_{d_1},\ldots,\tilde{\phi}_{d_{s-1}}]
  (x+i\frac{\gamma}{2})}
  {\text{W}_{\gamma}[\tilde{\phi}_{d_1},\ldots,\tilde{\phi}_{d_{s-1}}]
  (x-i\frac{\gamma}{2})}\,
  \frac{\text{W}_{\gamma}[\tilde{\phi}_{d_1},\ldots,\tilde{\phi}_{d_s}]
  (x-i\gamma)}
  {\text{W}_{\gamma}[\tilde{\phi}_{d_1},\ldots,\tilde{\phi}_{d_s}](x)},
  \label{Vhd1..ds}\\
  &\phi_{d_1\ldots d_s\,n}(x)\eqdef
  \hat{\mathcal{A}}_{d_1\ldots d_s}\phi_{d_1\ldots d_{s-1}\,n}(x)
  =\phi_{d_1\ldots d_s\,n}^*(x)\ \ (n\in\mathbb{Z}_{\geq 0})\n
  &\phantom{\phi_{d_1\ldots d_s\,n}(x)}=A(x)
  \text{W}_{\gamma}[\tilde{\phi}_{d_1},\ldots,\tilde{\phi}_{d_s},\phi_n](x),
  \label{phid1..dsn}\\
  &\qquad\ \,A(x)=
  \left(\frac{\sqrt{\prod_{j=0}^{s-1}V(x+i(\frac{s}{2}-j)\gamma)
  V^*(x-i(\frac{s}{2}-j)\gamma)}}
  {\text{W}_{\gamma}[\tilde{\phi}_{d_1},\ldots,\tilde{\phi}_{d_s}]
  (x-i\frac{\gamma}{2})
  \text{W}_{\gamma}[\tilde{\phi}_{d_1},\ldots,\tilde{\phi}_{d_s}]
  (x+i\frac{\gamma}{2})}\right)^{\frac12},\\
  &\tilde{\phi}_{d_1\ldots d_s\,\text{v}}(x)\eqdef
  \hat{\mathcal{A}}_{d_1\ldots d_s}
  \tilde{\phi}_{d_1\ldots d_{s-1}\,\text{v}}(x)
  =\tilde{\phi}_{d_1\ldots d_s\,\text{v}}^*(x)
  \ (\text{v}\in\{d_1,\ldots,d_M\}\backslash\{d_1,\ldots,d_s\})\n
  &\phantom{\tilde{\phi}_{d_1\ldots d_s\,\text{v}}(x)}=A(x)
  \text{W}_{\gamma}[\tilde{\phi}_{d_1},\ldots,\tilde{\phi}_{d_s},
  \tilde{\phi}_{\text{v}}](x),\\
  &\mathcal{H}_{d_1\ldots d_s}\phi_{d_1\ldots d_s\,n}(x)
  =\mathcal{E}_n\phi_{d_1\ldots d_s\,n}(x)
  \ \ (n\in\mathbb{Z}_{\geq 0}),
  \label{Hd1..dsphid1..dsn=}\\
  &\mathcal{H}_{d_1\ldots d_s}\tilde{\phi}_{d_1\ldots d_s\,\text{v}}(x)
  =\tilde{\mathcal{E}}_\text{v}\tilde{\phi}_{d_1\ldots d_s\,\text{v}}(x)
  \ \ (\text{v}\in\{d_1,\ldots,d_M\}\backslash\{d_1,\ldots,d_s\}),
  \label{Hd1..dsphid1..dsv=}\\
  &(\phi_{d_1\ldots d_s\,n},\phi_{d_1\ldots d_s\,m})
  =\prod_{j=1}^s(\mathcal{E}_n-\tilde{\mathcal{E}}_{d_j})\cdot
  h_n\delta_{nm}
  \ \ (n,m\in\mathbb{Z}_{\geq 0}),
  \label{(phid1..dsn,phid1..dsm)}
\end{align}
where $\text{W}_{\gamma}[f_1,\ldots,f_n]$ is the Casorati determinant of
a set of $n$ functions $\{f_j(x)\}$,
\begin{equation}
  \text{W}_{\gamma}[f_1,\ldots,f_n](x)
  \eqdef i^{\frac12n(n-1)}
  \det\Bigl(f_k\bigl(x^{(n)}_j\bigr)\Bigr)_{1\leq j,k\leq n},\quad
  x_j^{(n)}\eqdef x+i(\tfrac{n+1}{2}-j)\gamma,
  \label{Wdef}
\end{equation}
(for $n=0$, we set $\text{W}_{\gamma}[\cdot](x)=1$).
The operators $\hat{\mathcal{A}}_{d_1\ldots d_s}$ and
$\hat{\mathcal{A}}_{d_1\ldots d_s}^{\dagger}$ have no zero modes, which is the
characterization of virtual state wavefunctions (For cH case, we relax this
condition because these operators may be singular in the intermediate steps.).
Therefore the deformed systems are isospectral to the original system.
The deformed Hamiltonian $\mathcal{H}_{d_1\ldots d_s}$ can be
rewritten in the standard form:
\begin{align}
  &\mathcal{H}_{d_1\ldots d_s}
  =\mathcal{A}_{d_1\ldots d_s}^{\dagger}\mathcal{A}_{d_1\ldots d_s},
  \label{Hd1..dsstd}\\
  &\mathcal{A}_{d_1\ldots d_s}\eqdef
  i\bigl(e^{\frac{\gamma}{2}p}\sqrt{V_{d_1\ldots d_s}^*(x)}
  -e^{-\frac{\gamma}{2}p}\sqrt{V_{d_1\ldots d_s}(x)}\,\bigr),\n
  &\mathcal{A}_{d_1\ldots d_s}^{\dagger}\eqdef
  -i\bigl(\sqrt{V_{d_1\ldots d_s}(x)}\,e^{\frac{\gamma}{2}p}
  -\sqrt{V_{d_1\ldots d_s}^*(x)}\,e^{-\frac{\gamma}{2}p}\bigr),
  \label{Ad1..ds}\\
  &V_{d_1\ldots d_s}(x)\eqdef
  \sqrt{V(x-i\tfrac{s}{2}\gamma)V^*(x-i\tfrac{s+2}{2}\gamma)}\n
  &\phantom{V_{d_1\ldots d_s}(x)\eqdef}\times
  \frac{\text{W}_{\gamma}[\tilde{\phi}_{d_1},\ldots,\tilde{\phi}_{d_s}]
  (x+i\frac{\gamma}{2})}
  {\text{W}_{\gamma}[\tilde{\phi}_{d_1},\ldots,\tilde{\phi}_{d_s}]
  (x-i\frac{\gamma}{2})}\,
  \frac{\text{W}_{\gamma}[\tilde{\phi}_{d_1},\ldots,\tilde{\phi}_{d_s},
  \phi_0](x-i\gamma)}
  {\text{W}_{\gamma}[\tilde{\phi}_{d_1},\ldots,\tilde{\phi}_{d_s},\phi_0](x)}.
  \label{Vd1..ds}
\end{align}
These formulas \eqref{Hd1..ds}--\eqref{Vd1..ds} (except for
\eqref{(phid1..dsn,phid1..dsm)}) are derived algebraically
(\eqref{(phid1..dsn,phid1..dsm)} is derived only formally unless
$\mathcal{H}_{d_1\ldots d_{s'}}$ ($s'\leq s$) are hermitian).
The hermiticity of $\mathcal{H}_{d_1\ldots d_s}$ etc.\ should be considered in
each case.

\subsection{Multi-indexed orthogonal polynomials}
\label{sec:miop2}

We recapitulate the case-(1) multi-indexed orthogonal polynomials of cH, W and
AW types.

There are two types of the virtual state wavefunctions, type
$\I$ $\tilde{\phi}^{\I}_{\text{v}}(x)$ and type
$\II$ $\tilde{\phi}^{\II}_{\text{v}}(x)$ for cH, W and AW idQM systems.
The deformed systems are labeled by the index set $\mathcal{D}$,
\begin{align}
  &\mathcal{D}=\{d_1,\ldots,d_M\}\ \ (d_j\in\mathbb{Z}_{\geq 0}),\quad
  \mathcal{D}=\mathcal{D}^{\I}\,\cup\,\mathcal{D}^{\II},\quad
  M=M_{\I}+M_{\II},\n
  &\quad\mathcal{D}^{\I}\eqdef\{d\in\mathcal{D}\,|\,d:\text{type \I}\}
  =\{d^{\I}_1,\ldots,d^{\I}_{M_{\I}}\}
  \ \ (\text{$d^{\I}_j$ : mutually distinct}),
  \label{setD}\\
  &\quad\mathcal{D}^{\II}\eqdef\{d\in\mathcal{D}\,|\,d:\text{type \II}\}
  =\{d^{\II}_1,\ldots,d^{\II}_{M_{\II}}\}
  \ \ (\text{$d^{\II}_j$ : mutually distinct}),
  \nonumber
\end{align}
which are the degrees and types of the virtual state wavefunctions used in
$M$-step Darboux transformations.
The Hamiltonian is deformed as
$\mathcal{H}\to\mathcal{H}_{d_1}\to\mathcal{H}_{d_1d_2}\to\cdots\to
\mathcal{H}_{d_1\ldots d_s}\to\cdots\to
\mathcal{H}_{d_1\ldots d_M}=\mathcal{H}_{\mathcal{D}}$
by $M$-step Darboux transformations.
Various quantities of the deformed systems are denoted as
$\mathcal{H}_{\mathcal{D}}$, $\phi_{\mathcal{D}\,n}$,
$\mathcal{A}_{\mathcal{D}}$, etc.
Exactly speaking, $\mathcal{D}$ is an ordered set.
When the ordered set $\mathcal{D}$ is
$\mathcal{D}=\{d^{\I}_1,\ldots,d^{\I}_{M_{\I}},d^{\II}_1,\ldots,
d^{\II}_{M_{\II}}\}$ with $0\leq d^{\I}_1<\cdots<d^{\I}_{M_{\I}}$ and
$0\leq d^{\II}_1<\cdots<d^{\II}_{s_{\II}}$,
we call it the standard order.
Under the permutation of $d_j$'s, the deformed Hamiltonian
$\mathcal{H}_{\mathcal D}$ is invariant, but the denominator polynomial
$\check{\Xi}_{\mathcal{D}}(x)$ and the multi-indexed polynomials
$\check{P}_{\mathcal{D},n}(x)$ may change their signs.
Unless otherwise mentioned, we do not care much about the order of $\mathcal{D}$.

The denominator polynomial $\Xi_{\mathcal D}$ and the multi-indexed
polynomials $P_{\mathcal{D},n}$ are constructed as polynomial parts of the
Casoratians
$\text{W}_{\gamma}[\tilde{\phi}_{d_1},\ldots,\tilde{\phi}_{d_M}](x)$ and
$\text{W}_{\gamma}[\tilde{\phi}_{d_1},\ldots,\tilde{\phi}_{d_M},\phi_n](x)$,
respectively,
\begin{alignat}{2}
  \text{W}_{\gamma}[\tilde{\phi}_{d_1},\ldots,\tilde{\phi}_{d_M}](x)
  &=g_{\mathcal{D}}(x)\check{\Xi}_{\mathcal{D}}(x),&
  \check{\Xi}_{\mathcal{D}}(x)&\eqdef\Xi_{\mathcal{D}}\bigl(\eta(x)\bigr),
  \label{W=gDXiD}\\
  \text{W}_{\gamma}[\tilde{\phi}_{d_1},\ldots,\tilde{\phi}_{d_M},\phi_n](x)
  &=g^P_{\mathcal{D}}(x)\check{P}_{\mathcal{D},n}(x),&\quad
  \check{P}_{\mathcal{D},n}(x)&\eqdef P_{\mathcal{D},n}\bigl(\eta(x)\bigr),
  \label{W=gPDPDn}
\end{alignat}
whose concrete definitions are given by (3.18)--(3.19) of \cite{idQMcH} and
(3.37)--(3.38) of \cite{os27}, and
$g_{\mathcal{D}}(x)$ and $g^P_{\mathcal{D}}(x)$ can be read from
(3.25)--(3.26) of \cite{idQMcH} and
(3.50)--(3.51) of \cite{detmiop} with
$\text{W}_{\gamma}[\tilde{\phi}_{d_1},\ldots,\tilde{\phi}_{d_M}](x)
=\prod_{j=1}^M\phi_0(x^{(M)}_j)\cdot
\text{W}_{\gamma}[\nu^{-1}\check{\xi}_{d_1},\ldots,
\nu^{-1}\check{\xi}_{d_M}](x)$ and
$\text{W}_{\gamma}[\tilde{\phi}_{d_1},\ldots,\tilde{\phi}_{d_M},\phi_n](x)
$ $=\prod_{j=1}^{M+1}\phi_0(x^{(M+1)}_j)\cdot
\text{W}_{\gamma}[\nu^{-1}\check{\xi}_{d_1},\ldots,
\nu^{-1}\check{\xi}_{d_M},\check{P}_n](x)$
($\nu(x)$ and $\check{\xi}_{\text{v}}(x)$ are given in \cite{idQMcH,detmiop}).
We remark that $g_{\mathcal{D}}(x)$ and $g^P_{\mathcal{D}}(x)$ depend on
$M_{\I}$ and $M_{\II}$, but not on the specific values of $d_j$ and $n$.
Note that $\check{\Xi}^*_{\mathcal{D}}(x)=\check{\Xi}_{\mathcal{D}}(x)$ and
$\check{P}^*_{\mathcal{D},n}(x)=\check{P}_{\mathcal{D},n}(x)$.
The denominator polynomial $\Xi_{\mathcal{D}}(\eta)$ and the multi-indexed
polynomials $P_{\mathcal{D},n}(\eta)$ are polynomials in $\eta$ and their
degrees are $\ell_{\mathcal{D}}$ and $\ell_{\mathcal{D}}+n$,
respectively (we assume $c_{\mathcal{D}}^{\Xi}\neq 0$ and
$c_{\mathcal{D},n}^{P}\neq 0$, see (A.1)--(A.2) of \cite{idQMcH} and
(A.40)--(A.41) of \cite{equiv_miop}).
Here $\ell_{\mathcal{D}}$ is
\begin{equation}
  \ell_{\mathcal{D}}\eqdef\sum_{j=1}^Md_j-\tfrac12M(M-1)+2M_{\I}M_{\II}.
  \label{lD}
\end{equation}
We remark that $P_{\mathcal{D},0}$ and $\Xi_{\mathcal{D}}$ are proportional,
$\check{P}_{\mathcal{D},0}(x;\bm{\lambda})\propto
\check{\Xi}_{\mathcal{D}}(x;\bm{\lambda}+\bm{\delta})$,
(A.3) of \cite{idQMcH}, (3.44) of \cite{os27},
which is a consequence of the shape invariance.

The eigenfunctions of the deformed system \eqref{phid1..dsn} are expressed in
terms of the ground state wavefunction $\phi_0(x)$ with shifted parameters,
the denominator polynomials $\check{\Xi}_{\mathcal{D}}(x)$ and the multi-indexed
polynomial $\check{P}_{\mathcal{D},n}(x)$ as
\begin{align}
  \phi_{\mathcal{D}\,n}(x)&=c^{\phi}_{\mathcal{D}}\psi_{\mathcal{D}}(x)
  \check{P}_{\mathcal{D},n}(x)\quad(n\in\mathbb{Z}_{\geq 0}),
  \label{phiDn}\\
  \psi_{\mathcal{D}}(x;\bm{\lambda})&\eqdef
  \frac{\phi_0(x;\bm{\lambda}_{\mathcal{D}})}
  {\sqrt{\check{\Xi}_{\mathcal{D}}(x-i\frac{\gamma}{2};\bm{\lambda})
  \check{\Xi}_{\mathcal{D}}(x+i\frac{\gamma}{2};\bm{\lambda})}},
\end{align}
where $c^{\phi}_{\mathcal{D}}$ and $\bm{\lambda}_{\mathcal{D}}$
(which is denoted as $\bm{\lambda}^{[M_{\I},M_{\II}]}$ in previous papers) are
\begin{align}
  c^{\phi}_{\mathcal{D}}&=\alpha^{\I}(\bm{\lambda})^{\frac12M_{\I}}
  \alpha^{\II}(\bm{\lambda})^{\frac12M_{\II}}
  \kappa^{\frac14M(M-1)+M_{\I}M_{\II}},
  \label{cphiD}\\
  \bm{\lambda}_{\mathcal{D}}&\eqdef\bm{\lambda}+M_{\I}\tilde{\bm{\delta}}^{\I}
  +M_{\II}\tilde{\bm{\delta}}^{\II}.
  \label{lambdaD}
\end{align}
Here explicit forms of $\alpha(\bm{\lambda})$ and $\tilde{\bm{\delta}}$ are
given by (3.1), (3.3) of \cite{idQMcH} and
(3.25), (3.27) of \cite{os27},
and $\kappa$ is $1$ (for cH and W) and $q^{-1}$ (for AW). 

The deformed Hamiltonian $\mathcal{H}_{\mathcal{D}}$ in the standard form
\eqref{Hd1..dsstd} is specified by the potential function $V_{\mathcal{D}}$
\eqref{Vd1..ds} and it is expressed in terms of the potential function $V(x)$
with shifted parameters and the denominator polynomial
$\check{\Xi}_{\mathcal{D}}(x)$ (with shifted parameters),
\begin{equation}
  V_{\mathcal{D}}(x;\bm{\lambda})
  =V(x;\bm{\lambda}_{\mathcal{D}})\,
  \frac{\check{\Xi}_{\mathcal{D}}(x+i\frac{\gamma}{2};\bm{\lambda})}
  {\check{\Xi}_{\mathcal{D}}(x-i\frac{\gamma}{2};\bm{\lambda})}
  \frac{\check{\Xi}_{\mathcal{D}}(x-i\gamma;\bm{\lambda}+\bm{\delta})}
  {\check{\Xi}_{\mathcal{D}}(x;\bm{\lambda}+\bm{\delta})}.
  \label{VD}
\end{equation}
In order for the deformed Hamiltonian $\mathcal{H}_{\mathcal{D}}$ to be
hermitian, the parameters $\bm{\lambda}$ are restricted.
As a sufficient condition for the hermiticity, we have the following
\cite{os27,idQMcH}:
\begin{equation}
  \text{The denominator polynomial $\check{\Xi}_{\mathcal{D}}(x)$
  has no zero in $D_{\gamma}$ \eqref{Dgamma}.}
  \label{nozeroDg}
\end{equation}
For cH case, the degree of $\Xi_{\mathcal{D}}(\eta)$, $\ell_{\mathcal{D}}$,
should be even.
In the following, we assume that the range of parameters is chosen so that
the deformed Hamiltonian $\mathcal{H}_{\mathcal{D}}$ is hermitian.

The orthogonality of the eigenfunctions \eqref{(phid1..dsn,phid1..dsm)}, namely,
those of the multi-indexed polynomials $\check{P}_{\mathcal{D},n}(x)$ are
\begin{align}
  &\int_{x_1}^{x_2}\!\!dx\,\psi_{\mathcal{D}}(x)^2
  \check{P}_{\mathcal{D},n}(x)\check{P}_{\mathcal{D},m}(x)
  =h_{\mathcal{D},n}\delta_{nm}\quad(n,m\in\mathbb{Z}_{\geq 0}),
  \label{orthocPDn}\\
  &h_{\mathcal{D},n}=(c^{\phi}_{\mathcal{D}})^{-2}h_n
  \prod_{j=1}^{M_{\I}}(\mathcal{E}_n-\tilde{\mathcal{E}}^{\I}_{d^{\I}_j})\cdot
  \prod_{j=1}^{M_{\II}}(\mathcal{E}_n-\tilde{\mathcal{E}}^{\II}_{d^{\II}_j}),
  \label{hDn}
\end{align}
where explicit forms of $\tilde{\mathcal{E}}_{\text{v}}$ are given by
(3.7) of \cite{idQMcH} and (3.29) of \cite{os27}.
The multi-indexed orthogonal polynomial $P_{\mathcal{D},n}(\eta)$
has $n$ zeros in the physical region $\eta_{\text{min}}<\eta<\eta_{\text{max}}$
($\eta_{\text{min}}\eqdef\min(\eta(x_1),\eta(x_2))$,
$\eta_{\text{max}}\eqdef\max(\eta(x_1),\eta(x_2))$),
which interlace the $n+1$ zeros of $P_{\mathcal{D},n+1}(\eta)$ in the physical
region, and $\ell_{\mathcal{D}}$ zeros in the unphysical region
$\eta\in\mathbb{C}\backslash(\eta_{\text{min}},\eta_{\text{max}})$.

The similarity transformed Hamiltonian in terms of $\psi_{\mathcal{D}}(x)$
is square root free,
\begin{align}
  \widetilde{\mathcal{H}}_{\mathcal{D}}
  &\eqdef\psi_{\mathcal{D}}(x)^{-1}\circ
  \mathcal{H}_{\mathcal{D}}\circ\psi_{\mathcal{D}}(x)\n
  &=V(x;\bm{\lambda}_{\mathcal{D}})\,
  \frac{\check{\Xi}_{\mathcal{D}}(x+i\frac{\gamma}{2};\bm{\lambda})}
  {\check{\Xi}_{\mathcal{D}}(x-i\frac{\gamma}{2};\bm{\lambda})}
  \biggl(e^{\gamma p}
  -\frac{\check{\Xi}_{\mathcal{D}}(x-i\gamma;\bm{\lambda}+\bm{\delta})}
  {\check{\Xi}_{\mathcal{D}}(x;\bm{\lambda}+\bm{\delta})}\biggr)\n
  &\quad+V^*(x;\bm{\lambda}_{\mathcal{D}})\,
  \frac{\check{\Xi}_{\mathcal{D}}(x-i\frac{\gamma}{2};\bm{\lambda})}
  {\check{\Xi}_{\mathcal{D}}(x+i\frac{\gamma}{2};\bm{\lambda})}
  \biggl(e^{-\gamma p}
  -\frac{\check{\Xi}_{\mathcal{D}}(x+i\gamma;\bm{\lambda}+\bm{\delta})}
  {\check{\Xi}_{\mathcal{D}}(x;\bm{\lambda}+\bm{\delta})}\biggr),
  \label{tHD}
\end{align}
and the multi-indexed polynomials $\check{P}_{\mathcal{D},n}(x)$ are its
eigenpolynomials,
\begin{equation}
  \widetilde{\mathcal{H}}_{\mathcal{D}}\check{P}_{\mathcal{D},n}(x)
  =\mathcal{E}_n\check{P}_{\mathcal{D},n}(x)\quad(n\in\mathbb{Z}_{\geq 0}).
  \label{tHPDn=}
\end{equation}

\subsection{Some identities}
\label{sec:id}

For the ordered index set $\mathcal{D}=\{d_1,\ldots,d_M\}$, let us consider the
following ordered index sets,
\begin{equation}
  \mathcal{D}'=\{d_1,\ldots,d_M,d'\},\quad
  \mathcal{D}''=\{d_1,\ldots,d_M,d''\},\quad
  \mathcal{D}'''=\{d_1,\ldots,d_M,d',d''\},
  \label{D'D''D'''}
\end{equation}
where $d',d''\not\in\mathcal{D}$ and $d'\neq d''$.
{}From the properties of the multi-step Darboux transformations given in
\S\,\ref{sec:DT}, we have the following identities:
\begin{align}
  \phi_{\mathcal{D}''\,n}(x)&=\hat{\mathcal{A}}_{\mathcal{D}''}
  \frac{\hat{\mathcal{A}}_{\mathcal{D}'}^{\dagger}}
  {\mathcal{E}_n-\tilde{\mathcal{E}}_{d'}}
  \phi_{\mathcal{D}'\,n}(x),
  \label{idmaru1}\\
  \phi_{\mathcal{D}'''\,n}(x)&=\hat{\mathcal{A}}_{\mathcal{D}'''}
  \hat{\mathcal{A}}_{\mathcal{D}'}\phi_{\mathcal{D}\,n}(x).
  \label{idmaru2}
\end{align}
By extracting the polynomial parts from these identities, we obtain the
following proposition.
\begin{prop}\label{prop:id}
~\\
(1) When the types of $d'$ and $d''$ are the same, \eqref{idmaru1} gives
\begin{equation}
  (\mathcal{E}_n-\tilde{\mathcal{E}}_{d'})\biggl(
  \frac{\check{\Xi}_{\mathcal{D}'}(x-i\frac{\gamma}{2})}
  {\check{\Xi}_{\mathcal{D}''}(x-i\frac{\gamma}{2})}
  +\frac{\check{\Xi}_{\mathcal{D}'}(x+i\frac{\gamma}{2})}
  {\check{\Xi}_{\mathcal{D}''}(x+i\frac{\gamma}{2})}
  \biggr)\check{P}_{\mathcal{D}'',n}(x)
  =(\widetilde{\mathcal{H}}_{\mathcal{D}''}+\mathcal{E}_n
  -\tilde{\mathcal{E}}_{d'}-\tilde{\mathcal{E}}_{d''})
  \check{P}_{\mathcal{D}',n}(x).
  \label{idmaru1P}
\end{equation}
(2) When the types of $d'$ and $d''$ are different, \eqref{idmaru2} gives
\begin{equation}
  \kappa^{2M+\frac32}\sqrt{\alpha^{\Irm}\alpha^{\IIrm}}\,\biggl(
  \frac{\check{\Xi}_{\mathcal{D}}(x-i\frac{\gamma}{2})}
  {\check{\Xi}_{\mathcal{D}'''}(x-i\frac{\gamma}{2})}
  +\frac{\check{\Xi}_{\mathcal{D}}(x+i\frac{\gamma}{2})}
  {\check{\Xi}_{\mathcal{D}'''}(x+i\frac{\gamma}{2})}
  \biggr)\check{P}_{\mathcal{D}''',n}(x)
  =(\widetilde{\mathcal{H}}_{\mathcal{D}'''}+\mathcal{E}_n
  -\tilde{\mathcal{E}}_{d'}-\tilde{\mathcal{E}}_{d''})
  \check{P}_{\mathcal{D},n}(x).
  \label{idmaru2P}
\end{equation}
\end{prop}
Since the proof is rather technical, we present it in Appendix \ref{app:pr:id}.

\remark\label{rem:id}
The identities \eqref{idmaru1P} and \eqref{idmaru2P} are invariant
under the permutation of $d_j$, $d'$, $d''$ in $\mathcal{D}$, $\mathcal{D}'$,
$\mathcal{D}''$, $\mathcal{D}'''$, namely, the order of ordered sets
$\mathcal{D}$, $\mathcal{D}'$, $\mathcal{D}''$ and $\mathcal{D}'''$ are
irrelevant.

\remark\label{rem:commonzero}
For generic values of $\bm{\lambda}$, $\check{P}_{\mathcal{D},n}(x)$ and
$\check{\Xi}_{\mathcal{D}}(x\pm i\frac{\gamma}{2})$ do not have a common root.

\subsection{Discrete orthogonality relations}
\label{sec:domiop}

Let us fix a non-negative integer $\cN$ and set
$\cNt=\deg P_{\mathcal{D},\llcN}=\ell_{\mathcal{D}}+\cN$.
We denote the zeros of $P_{\mathcal{D},\llcN}(\eta)$ as
$\eta^{[\mathcal{D},\llcN]}_j$ ($j=1,2,\ldots,\cNt$), and assume that they
are simple.
We define $x^{[\mathcal{D},\llcN]}_j$ by requiring
$\eta(x^{[\mathcal{D},\llcN]}_j)=\eta^{[\mathcal{D},\llcN]}_j$ and
$x_1\leq\text{Re}\,x^{[\mathcal{D},\llcN]}_j\leq x_2$.

In the following, for simplicity in notation, we write
$\eta^{[\mathcal{D},\llcN]}_j$ and $x^{[\mathcal{D},\llcN]}_j$ as $\eta_j$
and $x_j$, respectively.
(Although this notation $x_j$ conflicts with the end points of the physical
range of the coordinate ($x_1$ and $x_2$), we think this does not cause any
confusion.)
Then $\check{P}_{\mathcal{D},\llcN}(x)$ is expressed as
\begin{equation}
  \check{P}_{\mathcal{D},\llcN}(x)
  =c^P_{\mathcal{D},\llcN}\prod_{j=1}^{\llcNt}\bigl(\eta(x)-\eta_j\bigr).
  \label{cPDNt}
\end{equation}

\subsubsection{polynomials $\cP_a$}
\label{sec:cPa}

For the index set $\mathcal{D}$ \eqref{setD} in the standard order,
let us define the index sets $\mathcal{D}'_{1,jk}$,
$\mathcal{D}'_{2,jk}$ and $\mathcal{D}'_{3,jk}$ as follows \cite{hs19}:
\begin{align}
  (1)\ \mathcal{D}'_{1,jk}&:
  \ E^{\I}\eqdef\{0,1,\ldots,d^{\I}_{M_{\I}}\}\backslash\mathcal{D}^{\I},\quad
  d^{\I}_j\in\mathcal{D}^{\I},\quad
  E^{\I}_j\eqdef\{\epsilon\in E^{\I}\,|\,\epsilon<d^{\I}_j\},\quad
  \epsilon^{\I}_k\in E^{\I}_j,\n
  &\ \ \ \mathcal{D}^{\prime\,\I}_{jk}\eqdef
  \bigl(\mathcal{D}^{\I}\backslash\{d^{\I}_j\}\bigr)\cup\{\epsilon^{\I}_k\},
  \quad\mathcal{D}'_{1,jk}\eqdef
  \mathcal{D}^{\prime\,\I}_{jk}\cup\mathcal{D}^{\II},
  \label{D'1jk}\\
  (2)\ \mathcal{D}'_{2,jk}&:
  \ E^{\II}\eqdef\{0,1,\ldots,d^{\II}_{M_{\II}}\}\backslash\mathcal{D}^{\II},
  \quad d^{\II}_j\in\mathcal{D}^{\II},\quad
  E^{\II}_j\eqdef\{\epsilon\in E^{\II}\,|\,\epsilon<d^{\II}_j\},\quad
  \epsilon^{\II}_k\in E^{\II}_j,\n
  &\ \ \ \mathcal{D}^{\prime\,\II}_{jk}\eqdef
  \bigl(\mathcal{D}^{\II}\backslash\{d^{\II}_j\}\bigr)\cup\{\epsilon^{\II}_k\},
  \quad\mathcal{D}'_{2,jk}\eqdef
  \mathcal{D}^{\I}\cup\mathcal{D}^{\prime\,\II}_{jk},
  \label{D'2jk}\\
  (3)\ \mathcal{D}'_{3,jk}&:
  \ d^{\I}_j\in\mathcal{D}^{\I},\quad d^{\II}_k\in\mathcal{D}^{\II},\quad
  \mathcal{D}'_{3,jk}\eqdef
  \bigl(\mathcal{D}^{\I}\backslash\{d^{\I}_j\}\bigr)\cup
  \bigl(\mathcal{D}^{\II}\backslash\{d^{\II}_k\}\bigr).
  \label{D'3jk}
\end{align}
As an ordered set, we choose one order of the elements for each
$\mathcal{D}'_{1,jk}$, $\mathcal{D}'_{2,jk}$ and $\mathcal{D}'_{3,jk}$,
e.g.\ the standard order.
Note that $|E^{\I}_j|=d^{\I}_j-(j-1)$ and $|E^{\II}_j|=d^{\II}_j-(j-1)$.
The numbers of these sets are
\begin{align}
  \#\{\mathcal{D}'_{1,jk}\}&=\sum_{j=1}^{M_{\I}}\bigl(d^{\I}_j-(j-1)\bigr)
  =\sum_{j=1}^{M_{\I}}d^{\I}_j-\frac12M_{\I}(M_{\I}-1),\n
  \#\{\mathcal{D}'_{2,jk}\}&=\sum_{j=1}^{M_{\II}}\bigl(d^{\II}_j-(j-1)\bigr)
  =\sum_{j=1}^{M_{\II}}d^{\II}_j-\frac12M_{\II}(M_{\II}-1),
  \label{kosuuD'}\\
  \#\{\mathcal{D}'_{3,jk}\}&=M_{\I}M_{\II},
  \nonumber
\end{align}
and the sum of these numbers is $\ell_{\mathcal{D}}$ \eqref{lD}.

Let us define the polynomials $\ccP_a(x)=\cP_a(\eta(x))$ ($a=1,\ldots,\cNt$)
as follows \cite{hs19}:
\begin{align}
  (0):&\ \ccP_a(x)=\check{P}_{\mathcal{D},n}(x)\ \ (0\leq n<\cN),\n
  (1):&\ \ccP_a(x)=\check{P}_{\mathcal{D}'_{1,jk},\llcN}(x)
  \ \ (\text{$\mathcal{D}'_{1,jk}$ in \eqref{D'1jk}}),\n
  (2):&\ \ccP_a(x)=\check{P}_{\mathcal{D}'_{2,jk},\llcN}(x)
  \ \ (\text{$\mathcal{D}'_{2,jk}$ in \eqref{D'2jk}}),
  \label{ccPadef}\\
  (3):&\ \ccP_a(x)=\check{P}_{\mathcal{D}'_{3,jk},\llcN}(x)
  \ \ (\text{$\mathcal{D}'_{3,jk}$ in \eqref{D'3jk}}).
  \nonumber
\end{align}
The total number of these $\ccP_a(x)$ is actually $\cN+\ell_{\mathcal{D}}=\cNt$.

Let us show that the conditions in \eqref{cPacond} are satisfied.
The condition (\romannumeral1) is trivial.
The condition (\romannumeral2) is satisfied, because we have
\begin{align}
  (0):&\ \deg\cP_a=\ell_{\mathcal{D}}+n,\ \ n<\cN,\n
  (1):&\ \deg\cP_a=\ell_{\mathcal{D}'_{1,jk}}+\cN,
  \ \ \ell_{\mathcal{D}'_{1,jk}}=\ell_{\mathcal{D}}-d^{\I}_j+\epsilon^{\I}_k
  <\ell_{\mathcal{D}},\n
  (2):&\ \deg\cP_a=\ell_{\mathcal{D}'_{2,jk}}+\cN,
  \ \ \ell_{\mathcal{D}'_{2,jk}}=\ell_{\mathcal{D}}-d^{\II}_j+\epsilon^{\II}_k
  <\ell_{\mathcal{D}},
  \label{cPdeg}\\
  (3):&\ \deg\cP_a=\ell_{\mathcal{D}'_{3,jk}}+\cN,
  \ \ \ell_{\mathcal{D}'_{3,jk}}=\ell_{\mathcal{D}}-d^{\I}_j-d^{\II}_k-1
  <\ell_{\mathcal{D}}.
  \nonumber
\end{align}
The condition (\romannumeral3) is satisfied by the following proposition.
\begin{prop}\label{prop:HDtccPa}
For $\ccP_a(x)$ \eqref{ccPadef}, we have
\begin{equation}
  \widetilde{\mathcal{H}}_{\mathcal{D}}\ccP_a(x)\bigl|_{x=x_j}\,=
  \mathcal{E}^{\cP}_a\cP_a(\eta_j),
  \label{HDtccPa}
\end{equation}
where $\mathcal{E}^{\cP}_a$ are given by
\begin{align}
  (0):&\ \mathcal{E}^{\cP}_a=\mathcal{E}_n,\n
  (1):&\ \mathcal{E}^{\cP}_a=\tilde{\mathcal{E}}^{\Irm}_{d^{\Irm}_j}
  +\tilde{\mathcal{E}}^{\Irm}_{\epsilon^{\Irm}_k}-\mathcal{E}_{\llcN},\n
  (2):&\ \mathcal{E}^{\cP}_a=\tilde{\mathcal{E}}^{\IIrm}_{d^{\IIrm}_j}
  +\tilde{\mathcal{E}}^{\IIrm}_{\epsilon^{\IIrm}_k}-\mathcal{E}_{\llcN},
  \label{EcPa}\\
  (3):&\ \mathcal{E}^{\cP}_a=\tilde{\mathcal{E}}^{\Irm}_{d^{\Irm}_j}
  +\tilde{\mathcal{E}}^{\IIrm}_{d^{\IIrm}_k}-\mathcal{E}_{\llcN}.
  \nonumber
\end{align}
\end{prop}
Proof:\\
(0): Eq.\,\eqref{tHPDn=} gives
$\widetilde{\mathcal{H}}_{\mathcal{D}}\ccP_a(x)=\mathcal{E}_n\ccP_a(x)$,
and \eqref{HDtccPa} is obtained by setting $x=x_j$.\\
(1): Eq.\,\eqref{idmaru1P} with the replacement
$(\mathcal{D}',\mathcal{D}'',n,d',d'')\to
(\mathcal{D}'_{1,jk},\mathcal{D},\cN,\epsilon^{\I}_k,d^{\I}_j)$ gives
\begin{equation*}
  (\cdots)\times\check{P}_{\mathcal{D},\llcN}(x)
  =(\widetilde{\mathcal{H}}_{\mathcal{D}}+\mathcal{E}_{\llcN}
  -\tilde{\mathcal{E}}^{\I}_{\epsilon^{\I}_k}
  -\tilde{\mathcal{E}}^{\I}_{d^{\I}_j})\ccP_a(x),
\end{equation*}
and \eqref{HDtccPa} is obtained by setting $x=x_j$.\\
(2): Eq.\,\eqref{idmaru1P} with the replacement
$(\mathcal{D}',\mathcal{D}'',n,d',d'')\to
(\mathcal{D}'_{2,jk},\mathcal{D},\cN,\epsilon^{\II}_k,d^{\II}_j)$ gives
\begin{equation*}
  (\cdots)\times\check{P}_{\mathcal{D},\llcN}(x)
  =(\widetilde{\mathcal{H}}_{\mathcal{D}}+\mathcal{E}_{\llcN}
  -\tilde{\mathcal{E}}^{\II}_{\epsilon^{\II}_k}
  -\tilde{\mathcal{E}}^{\II}_{d^{\II}_j})\ccP_a(x),
\end{equation*}
and \eqref{HDtccPa} is obtained by setting $x=x_j$.\\
(3): Eq.\,\eqref{idmaru2P} with the replacement
$(\mathcal{D},\mathcal{D}''',n,d',d'')\to
(\mathcal{D}'_{3,jk},\mathcal{D},\cN,d^{\I}_j,d^{\II}_k)$ gives
\begin{equation*}
  (\cdots)\times\check{P}_{\mathcal{D},\llcN}(x)
  =(\widetilde{\mathcal{H}}_{\mathcal{D}}+\mathcal{E}_{\llcN}
  -\tilde{\mathcal{E}}^{\I}_{d^{\I}_j}
  -\tilde{\mathcal{E}}^{\II}_{d^{\II}_k})\ccP_a(x),
\end{equation*}
and \eqref{HDtccPa} is obtained by setting $x=x_j$.
\hfill\fbox{}

\medskip
\noindent
By using \eqref{EcPa} and explicit forms of $\mathcal{E}_n$ and
$\tilde{\mathcal{E}}_{\text{v}}$, we can show that the condition
(\romannumeral4) is satisfied for generic values of $\bm{\lambda}$.
Thus the polynomials $\ccP_a(x)$ \eqref{ccPadef} (with generic values of
$\bm{\lambda}$) satisfy all the conditions in \eqref{cPacond}.

\subsubsection{matrices $\cMt$ and $\cM$}
\label{sec:cM}

Let us show the proposition for $\cMt$.
\begin{prop}\label{prop:cMtjk}
The matrix elements $\cMt_{jk}$ \eqref{cMtdef} for $j\neq k$ are expressed as
\begin{equation}
  \cMt_{jk}=\frac{1}{\bigl(\eta(x_j-i\gamma)-\eta_k\bigr)
  \bigl(\eta(x_j+i\gamma)-\eta_k\bigr)}
  \frac{-\widetilde{\mathcal{H}}_{\mathcal{D}}
  \bigl(\eta(x)\check{P}_{\mathcal{D},\llcN}(x)\bigr)\bigl|_{x=x_j}}
  {P'_{\mathcal{D},\llcN}(\eta_j)}\quad(j\neq k).
  \label{cMtjk1}
\end{equation}
\end{prop}
Proof: The similarity transformed Hamiltonian
$\widetilde{\mathcal{H}}_{\mathcal{D}}$ \eqref{tHD} has the following form,
\begin{equation*}
  \widetilde{\mathcal{H}}_{\mathcal{D}}
  =A(x)\bigl(e^{\gamma p}-B(x)\bigr)+A^*(x)\bigl(e^{-\gamma p}-B^*(x)\bigr).
\end{equation*}
Recalling \eqref{cPDNt} and evaluating
$\widetilde{\mathcal{H}}_{\mathcal{D}}\check{P}_{\mathcal{D},\llcN}(x)
=\mathcal{E}_{\llcN}\check{P}_{\mathcal{D},\llcN}(x)$ at $x=x_j$, we have
\begin{equation}
  A(x_j)\prod_{l=1}^{\llcNt}\bigl(\eta(x_j-i\gamma)-\eta_l\bigr)
  +A^*(x_j)\prod_{l=1}^{\llcNt}\bigl(\eta(x_j+i\gamma)-\eta_l\bigr)=0.
  \label{Axj+A*xj}
\end{equation}
For $j\neq k$, we have
\begin{align*}
  &\quad\widetilde{\mathcal{H}}_{\mathcal{D}}
  \prod_{\genfrac{}{}{0pt}{}{l=1}{l\neq k}}^{\llcNt}
  \bigl(\eta(x)-\eta_l\bigr)\Bigl|_{x=x_j}\\
  &=A(x_j)\prod_{\genfrac{}{}{0pt}{}{l=1}{l\neq k}}^{\llcNt}
  \bigl(\eta(x_j-i\gamma)-\eta_l\bigr)
  +A^*(x_j)\prod_{\genfrac{}{}{0pt}{}{l=1}{l\neq k}}^{\llcNt}
  \bigl(\eta(x_j+i\gamma)-\eta_l\bigr)\\
  &=\frac{A(x_j)\prod\limits_{l=1}^{\llcNt}
  \bigl(\eta(x_j-i\gamma)-\eta_l\bigr)}{\eta(x_j-i\gamma)-\eta_k}
  +\frac{A^*(x_j)\prod\limits_{l=1}^{\llcNt}
  \bigl(\eta(x_j+i\gamma)-\eta_l\bigr)}{\eta(x_j+i\gamma)-\eta_k}\\
  &\stackrel{(\text{\romannumeral1})}{=}
  \frac{A(x_j)\prod\limits_{l=1}^{\llcNt}
  \bigl(\eta(x_j-i\gamma)-\eta_l\bigr)\cdot
  \bigl(\eta(x_j+i\gamma)-\eta(x_j-i\gamma)\bigr)}
  {\bigl(\eta(x_j-i\gamma)-\eta_k\bigr)\bigl(\eta(x_j+i\gamma)-\eta_k\bigr)}\\
  &\stackrel{(\text{\romannumeral1})}{=}
  \frac{-A(x_j)\eta(x_j-i\gamma)\prod\limits_{l=1}^{\llcNt}
  \bigl(\eta(x_j-i\gamma)-\eta_l\bigr)
  -A^*(x_j)\eta(x_j+i\gamma)\prod\limits_{l=1}^{\llcNt}
  \bigl(\eta(x_j+i\gamma)-\eta_l\bigr)}
  {\bigl(\eta(x_j-i\gamma)-\eta_k\bigr)\bigl(\eta(x_j+i\gamma)-\eta_k\bigr)}\\
  &=\frac{-\widetilde{\mathcal{H}}_{\mathcal{D}}\Bigl(\eta(x)
  \prod\limits_{l=1}^{\llcNt}\bigl(\eta(x)-\eta_l\bigr)\Bigr)\Bigl|_{x=x_j}}
  {\bigl(\eta(x_j-i\gamma)-\eta_k\bigr)\bigl(\eta(x_j+i\gamma)-\eta_k\bigr)},
\end{align*}
where we have used \eqref{Axj+A*xj} in (\romannumeral1).
Then the matrix elements $\cMt_{jk}$ \eqref{cMtdef} for $j\neq k$ become
\begin{equation*}
  \cMt_{jk}=\frac{1}{\bigl(\eta(x_j-i\gamma)-\eta_k\bigr)
  \bigl(\eta(x_j+i\gamma)-\eta_k\bigr)}
  \frac{-\widetilde{\mathcal{H}}_{\mathcal{D}}\Bigl(\eta(x)
  \prod\limits_{l=1}^{\llcNt}\bigl(\eta(x)-\eta_l\bigr)\Bigr)\Bigl|_{x=x_j}}
  {\prod\limits_{\genfrac{}{}{0pt}{}{l=1}{l\neq j}}^{\llcNt}(\eta_j-\eta_l)}.
\end{equation*}
By multiplying the numerator and denominator of the last factor by
$c^P_{\mathcal{D},\llcN}$, \eqref{cMtjk1} is obtained.
\hfill\fbox{}

\medskip
Let us define $F_j$ ($j=1,\ldots,\cNt$) as
\begin{equation}
  F_j\eqdef-\frac{\widetilde{\mathcal{H}}_{\mathcal{D}}
  \bigl(\eta(x)\check{P}_{\mathcal{D},\llcN}(x)\bigr)\bigl|_{x=x_j}}
  {P'_{\mathcal{D},\llcN}(\eta_j)}.
  \label{Fj}
\end{equation}
Then we have the following proposition.
\begin{prop}\label{prop:Fj}
The number $F_j$ \eqref{Fj} is expressed as 
\begin{equation}
  F_j=\check{F}(x_j)=F(\eta_j).
  \label{FjFxj}
\end{equation}
Here $\check{F}(x)$ is given by
\begin{align}
  \check{F}(x)\eqdefrm\frac{-1}{P'_{\mathcal{D},\llcN}(\eta(x))}
  &\biggl(\eta(x-i\gamma)V(x;\bm{\lambda}_{\mathcal{D}})\,
  \frac{\check{\Xi}_{\mathcal{D}}(x+i\frac{\gamma}{2})}
  {\check{\Xi}_{\mathcal{D}}(x-i\frac{\gamma}{2})}
  \check{P}_{\mathcal{D},\llcN}(x-i\gamma)\n
  &+\eta(x+i\gamma)V^*(x;\bm{\lambda}_{\mathcal{D}})\,
  \frac{\check{\Xi}_{\mathcal{D}}(x-i\frac{\gamma}{2})}
  {\check{\Xi}_{\mathcal{D}}(x+i\frac{\gamma}{2})}
  \check{P}_{\mathcal{D},\llcN}(x+i\gamma)\biggr).
  \label{cFx}
\end{align}
It is a rational function of $\eta(x)$, and $F(\eta)$ is given by
$F(\eta(x))\eqdefrm\check{F}(x)$.
\end{prop}
Proof: By \eqref{tHD}, $F_j$ becomes
\begin{align*}
  F_j=\frac{-1}{P'_{\mathcal{D},\llcN}(\eta_j)}
  &\biggl(V(x_j;\bm{\lambda}_{\mathcal{D}})\,
  \frac{\check{\Xi}_{\mathcal{D}}(x_j+i\frac{\gamma}{2})}
  {\check{\Xi}_{\mathcal{D}}(x_j-i\frac{\gamma}{2})}
  \eta(x_j-i\gamma)\check{P}_{\mathcal{D},\llcN}(x_j-i\gamma)\\
  &+V^*(x_j;\bm{\lambda}_{\mathcal{D}})\,
  \frac{\check{\Xi}_{\mathcal{D}}(x_j-i\frac{\gamma}{2})}
  {\check{\Xi}_{\mathcal{D}}(x_j+i\frac{\gamma}{2})}
  \eta(x_j+i\gamma)\check{P}_{\mathcal{D},\llcN}(x_j+i\gamma)\biggr),
\end{align*}
namely $F_j=\check{F}(x_j)$.
Let us show that $\check{F}(x)$ is a rational function of $\eta(x)=x,x^2$ and
$\cos x$ for cH, W and AW cases, respectively. From \eqref{cFx} and explicit
form of $V(x)$, $\check{F}(x)$ is a rational function of $x$ (cH, W) or
$e^{ix}$ (AW).
For cH case, it is trivial that $\check{F}(x)$ is a rational function of $x$.
For W and AW cases, the potential function $V(x)$ satisfies
\begin{equation}
  \text{W,\,AW}\ :\ V^*(x)=V(-x).
  \label{V-x}
\end{equation}
By using this and $\eta(-x)=\eta(x)$, we obtain $\check{F}(-x)=\check{F}(x)$.
This means that $\check{F}(x)$ is a rational function of $x^2$ (W) or
$e^{ix}+e^{-ix}=2\cos x$ (AW).
\hfill\fbox{}

\medskip
\remark\label{rem:Fj}
By \eqref{Axj+A*xj}, $F_j$ can be written as
\begin{equation}
  F_j=\frac{\eta(x_j+i\gamma)-\eta(x_j-i\gamma)}
  {P'_{\mathcal{D},\llcN}(\eta_j)}
  V(x_j;\bm{\lambda}_{\mathcal{D}})\,
  \frac{\check{\Xi}_{\mathcal{D}}(x_j+i\frac{\gamma}{2})}
  {\check{\Xi}_{\mathcal{D}}(x_j-i\frac{\gamma}{2})}
  \check{P}_{\mathcal{D},\llcN}(x_j-i\gamma).
  \label{Fj2}
\end{equation}

\medskip
The sinusoidal coordinates have the following property.
\begin{lemma}\label{lem:eta}
 The sinusoidal coordinates $\eta(x)=x,x^2$ and $\cos x$ satisfy the following
identity for any complex numbers $a,b$ and $c$:
\begin{equation}
  \bigl(\eta(a-c)-\eta(b)\bigr)\bigl(\eta(a+c)-\eta(b)\bigr)
  =\bigl(\eta(b-c)-\eta(a)\bigr)\bigl(\eta(b+c)-\eta(a)\bigr).
  \label{etaid}
\end{equation}
\end{lemma}
Proof: Direct calculation shows this lemma.
\hfill\fbox{}

\medskip
\noindent
By the similarity transformation \eqref{cM}, we obtain the symmetric matrix
$\cM$.
\begin{prop}\label{prop:cM}
By taking $g_j=\sqrt{F_j}$ in \eqref{cM}, the matrix $\cM$ is symmetric,
\eqref{cMsym}. 
\end{prop}
Proof: Since the matrix elements $\cMt_{jk}$ ($j\neq k$) are expressed as
\begin{equation}
  \cMt_{jk}=\frac{F_j}{\bigl(\eta(x_j-i\gamma)-\eta_k\bigr)
  \bigl(\eta(x_j+i\gamma)-\eta_k\bigr)}\quad(j\neq k),
  \label{cMtjk}
\end{equation}
the matrix elements $\cM_{jk}$ ($j\neq k$) become
\begin{equation}
  \cM_{jk}=\frac{\sqrt{F_j}\,\sqrt{F_k}}{\bigl(\eta(x_j-i\gamma)-\eta_k\bigr)
  \bigl(\eta(x_j+i\gamma)-\eta_k\bigr)}\quad(j\neq k).
  \label{cMjk}
\end{equation}
By Lemma\,\ref{lem:eta}, $\cM_{jk}$ are symmetric in $j$ and $k$.
\hfill\fbox{}

\medskip
\remark\label{rem:cMtjj}
We have not written down the diagonal elements $\cM_{jj}(=\cMt_{jj})$
explicitly, because their concrete forms are not needed to show \eqref{cMsym}.
For the sake of completeness, we write down the concrete form of
$\cMt_{jj}=\cM_{jj}$,
\begin{align}
  \cMt_{jj}&=\frac{F_j}{\bigl(\eta(x_j-i\gamma)-\eta_j\bigr)
  \bigl(\eta(x_j+i\gamma)-\eta_j\bigr)}\n
  &\quad-V(x_j;\bm{\lambda}_{\mathcal{D}})\,
  \frac{\check{\Xi}_{\mathcal{D}}(x_j+i\frac{\gamma}{2};\bm{\lambda})}
  {\check{\Xi}_{\mathcal{D}}(x_j-i\frac{\gamma}{2};\bm{\lambda})}
  \frac{\check{\Xi}_{\mathcal{D}}(x_j-i\gamma;\bm{\lambda}+\bm{\delta})}
  {\check{\Xi}_{\mathcal{D}}(x_j;\bm{\lambda}+\bm{\delta})}\n
  &\quad-V^*(x_j;\bm{\lambda}_{\mathcal{D}})\,
  \frac{\check{\Xi}_{\mathcal{D}}(x_j-i\frac{\gamma}{2};\bm{\lambda})}
  {\check{\Xi}_{\mathcal{D}}(x_j+i\frac{\gamma}{2};\bm{\lambda})}
  \frac{\check{\Xi}_{\mathcal{D}}(x_j+i\gamma;\bm{\lambda}+\bm{\delta})}
  {\check{\Xi}_{\mathcal{D}}(x_j;\bm{\lambda}+\bm{\delta})},
  \label{cMtjj}
\end{align}
which is derived from the definition \eqref{cMtdef} by using \eqref{Axj+A*xj},
like as \eqref{cMtjk1} and \eqref{cFx}.

\subsubsection{discrete orthogonality relations}
\label{sec:domiop2}

Let us present a main result of this paper.
\begin{thm}\label{thm:domiop}
For the case-(1) multi-indexed orthogonal polynomials of cH, W and AW types
$P_{\mathcal{D},n}(\eta)$, $\cP_a(\eta)$ \eqref{ccPadef} and $F(\eta)$
\eqref{FjFxj}, we have the discrete orthogonality relations:
\begin{equation}
  \sum_{j=1}^{\cNt}\frac{1}{F(\eta_j)}
  \frac{\cP_a(\eta_j)}{P'_{\mathcal{D},\llcN}(\eta_j)}
  \frac{\cP_b(\eta_j)}{P'_{\mathcal{D},\llcN}(\eta_j)}
  =k^{[\mathcal{D},\llcN]}_a\delta_{ab}\quad(a,b=1,2,\ldots,\cNt),
  \label{domiop}
\end{equation}
where $k^{[\mathcal{D},\llcN]}_a$'s are normalization constants.
\end{thm}
Proof: Since the assumptions of Theorem\,\ref{thm:do} are satisfied (for generic
values of $\bm{\lambda}$), Theorem\,\ref{thm:do} gives this theorem.
\hfill\fbox{}

\medskip
\remark\label{rem:Ddep}
In contrast to the weight $1/c_2(\eta)$ in oQM case, this weight $1/F(\eta)$
depends on the multi index $\mathcal{D}$, see \eqref{cFx}.

\remark\label{rem:posi2}
As in Remark\,\ref{rem:posi}, the weights $1/F(\eta_j)$ may not be
positive and may be complex, and the normalization constants
$k^{[\mathcal{D},\llcN]}_a$ may not be positive.
As in Remark\,\ref{rem:kNa}, the theorem states nothing about the properties
of $k^{[\mathcal{D},\llcN]}_a$, nor does it state anything for $\cNt=0,1$ cases.

\remark\label{rem:HSVt}
The property $P^*_{\mathcal{D},\llcN}(\eta)=P_{\mathcal{D},\llcN}(\eta)$
(which holds if \eqref{condcH}--\eqref{condWAW} are satisfied) implies
\begin{equation}
  \{\eta_1^*,\ldots,\eta_{\llcNt}^*\}=\{\eta_1,\ldots,\eta_{\llcNt}\}
  \ \ (\text{as a set}).
  \label{etaj*}
\end{equation}
Base on this fact, let us define $\bar{j}$ as $\eta_{\bar{j}}\eqdef\eta_j^*$.
If $\eta_j^*=\eta_j$, we have $\bar{j}=j$.
In \cite{hs19}, the vector space
$\widetilde{\bm{V}}=\{\tilde{v}=(\tilde{v}_j)\in\mathbb{C}^{\llcNt}\,|\,
\tilde{v}_j^*=\tilde{v}_{\bar{j}}\}$ is considered.
It is a $\cNt$-dimensional vector space over $\mathbb{R}$, and
$(\tilde{v},\tilde{w})=\sum_{j=1}^{\llcNt}\tilde{v}_j\tilde{w}_j$ gives an
(indefinite) inner product on $\widetilde{\bm{V}}$.
By studying $(\cMt^{\text{HS}}_{jk})^*$, it is shown that $\widetilde{\bm{V}}$
is invariant under the action of $\cMt^{\text{HS}}$ \cite{hs19}.
A similar analysis is possible for our $\cMt$.

\remark\label{rem:anypara}
The discrete orthogonality relations \eqref{domiop} are shown for the
multi-indexed orthogonal polynomials $P_{\mathcal{D},n}$.
As mentioned in Remark\,\ref{rem:kNa}, the difference equations \eqref{tHPDn=}
are used in the proof, but the orthogonality relations \eqref{orthocPDn} are
not explicitly used.
In oder for $P_{\mathcal{D},n}$ to be orthogonal polynomials, the parameters
$\bm{\lambda}$ are restricted by the conditions such as \eqref{nozeroDg}.
If the parameters $\bm{\lambda}$ do not satisfy the conditions, the polynomials
$P_{\mathcal{D},n}$ are no longer orthogonal polynomials, but may still satisfy
the difference equations. Then, the discrete orthogonality relations still
hold in that case. Let us discuss this point.
For the parameters $\bm{\lambda}$ satisfying \eqref{condcH}--\eqref{condWAW},
they have 4 real degrees of freedom, and one more degree of freedom $q$
($0<q<1$) for AW case.
Let us consider $a_1,a_2,a_3,a_4,q\in\mathbb{C}$ ($a_i$ is $\lambda_i$ or
$q^{\lambda_i}$) without any restriction (except for the condition
$c^{\Xi}_{\mathcal{D}},c^P_{\mathcal{D},\llcN}\neq 0$).
There are 8 real degrees of freedom for cH and W, and 10 for AW.
The definitions of the polynomials $P_{\mathcal{D},n}$,
(3.19) of \cite{idQMcH} and (3.38) of \cite{os27},
are meaningful for these complex values of parameters.
Note that the property $P^*_{\mathcal{D},n}=P_{\mathcal{D},n}$ is lost and
\eqref{etaj*} does not hold.
The difference equations \eqref{tHPDn=} hold by replacing
$V^*(x;\bm{\lambda}_{\mathcal{D}})$ in \eqref{tHD} as follows:
\begin{equation}
  V^*(x)\ \to\ \left\{
  \begin{array}{ll}
  (a_3-ix)(a_4-ix)&:\text{cH}\\[2pt]
  V(-x)&:\text{W,\,AW}
  \end{array}\right..
\end{equation}
Then Proposition\,\ref{prop:id}, the conditions \eqref{cPacond} for $\ccP_a(x)$
\eqref{ccPadef}, Proposition\,\ref{prop:cMtjk}--\ref{prop:cM} are valid.
Thus the discrete orthogonality relations \eqref{domiop} hold for any complex
values of parameters (with the condition
$c^{\Xi}_{\mathcal{D}},c^P_{\mathcal{D},\llcN},F(\eta_j)\neq0$).
We can verify this for small $M$, $d_j$, $\cN$ and $n$ by numerical calculation.
The normalization constants $k^{[\mathcal{D},\llcN]}_a$ may not be real.

\remark\label{rem:MP}
The case-(1) multi-indexed orthogonal polynomials are also constructed for the
Meixner-Pollaczek (MP) polynomials \cite{idQMcH}.
In this case, there is only one type of virtual states.
Various formulas are obtained from those of two type case (cH,W,AW) given above,
by setting $M_{\II}=0$ and neglecting a superscript (or subscript) $\I$.
There is no $D'_{2,jk}$ and $D'_{3,jk}$. So, the polynomials $\ccP_a(x)$
\eqref{ccPadef} are only (0) and (1).
The condition (\romannumeral4) in \eqref{cPacond} may not be satisfied even for
generic values of $\bm{\lambda}$. However, the case-(1) multi-indexed MP
polynomials $P_{\mathcal{D},n}(\eta)$ can be obtained from the case-(1)
multi-indexed cH polynomials by taking certain limit of parameters (with
appropriate rescalings).
By this limit, the discrete orthogonality relations \eqref{domiop} for the
case-(1) multi-indexed cH polynomials are inherited by the case-(1)
multi-indexed MP polynomials.
As in Remark\,\ref{rem:anypara}, they hold for any complex values of parameters.

\medskip
We conjecture the normalization constants $k^{[\mathcal{D},\llcN]}_a$ as follows.
\begin{conj}\label{conj:kDNa}
The normalization constants $k^{[\mathcal{D},\llcN]}_a$ are given by
\begin{align}
  &(0)\ \cP_a=P_{\mathcal{D},n}:\ k^{[\mathcal{D},\llcN]}_a
  =\frac{h_{\mathcal{D},n}}{h_{\mathcal{D},\llcN}}\times\left\{
  \begin{array}{ll}
  {\displaystyle\frac{1}{2(b_1+2\cN-1)}}&:\text{\rm cH,\,W}\\[10pt]
  {\displaystyle\frac{q^{\cN+1}}{(1-q^2)(1-b_4q^{2\cN-1})}}&:\text{\rm AW}
  \end{array}\right.,\\
  &(1)\ \cP_a=P_{\mathcal{D}'_{1,jk},\llcN}:\n
  &\phantom{(1)\ }\text{\rm cH}:\ k^{[\mathcal{D},\llcN]}_a=
  \frac{h_{\mathcal{D}'_{1,jk},\llcN}}{h_{\mathcal{D},\llcN}}\,
  \frac{(\epsilon^{\Irm}_k+1)_{d^{\Irm}_j-\epsilon^{\Irm}_k}}{2}\,
  \frac{1}{(a_1+a_3-d^{\Irm}_j-1,a_2+a_4+\epsilon^{\Irm}_k
  )_{d^{\Irm}_j-\epsilon^{\Irm}_k}}\n
  &\phantom{\phantom{(1)\ }\text{\rm cH}:\ k^{[\mathcal{D},\llcN]}_a=}
  \times\frac{1}{(a_1-a_2-d^{\Irm}_j)_{d^{\Irm}_j-\epsilon^{\Irm}_k}
  (a_3-a_4-d^{\Irm}_j)_{d^{\Irm}_j-\epsilon^{\Irm}_k}}\,
  \frac{(b^{\prime\,\text{\rm cH}}-d^{\Irm}_k)_{d^{\Irm}_j-\epsilon^{\Irm}_k}}
  {-b^{\prime\,\text{\rm cH}}+1+2\epsilon^{\Irm}_k}\n
  &\phantom{\phantom{(1)\ }\text{\rm cH}:\ k^{[\mathcal{D},\llcN]}_a=}
  \times\prod_{\genfrac{}{}{0pt}{}{i=1}{i\neq j}}^{M_{\Irm}}
  \frac{d^{\Irm}_i-\epsilon^{\Irm}_k}{d^{\Irm}_i-d^{\Irm}_j}\,
  \frac{-b^{\prime\,\text{\rm cH}}+d^{\Irm}_i+\epsilon^{\Irm}_k+1}
  {-b^{\prime\,\text{\rm cH}}+d^{\Irm}_i+d^{\Irm}_j+1}\cdot
  \prod_{i=1}^{M_{\IIrm}}
  \frac{d^{\IIrm}_i+\epsilon^{\Irm}_k+1}{d^{\IIrm}_i+d^{\Irm}_j+1}\,
  \frac{b^{\prime\,\text{\rm cH}}+d^{\IIrm}_i-\epsilon^{\Irm}_k}
  {b^{\prime\,\text{\rm cH}}+d^{\IIrm}_i-d^{\Irm}_j},\n
  &\phantom{(1)\ }\ \text{\rm W}:\ k^{[\mathcal{D},\llcN]}_a=
  \frac{h_{\mathcal{D}'_{1,jk},\llcN}}{h_{\mathcal{D},\llcN}}\,
  \frac{1}{2(\epsilon^{\Irm}_k+1)_{d^{\Irm}_j-\epsilon^{\Irm}_k}}\,
  \frac{1}{(a_1+a_2-d^{\Irm}_j-1,a_3+a_4+\epsilon^{\Irm}_k
  )_{d^{\Irm}_j-\epsilon^{\Irm}_k}}\n
  &\phantom{\phantom{(1)\ }\text{\rm W}:\ k^{[\mathcal{D},\llcN]}_a=}
  \times\prod_{l=1}^2\prod_{m=3}^4
  \frac{1}{(a_l-a_m-d^{\Irm}_j)_{d^{\Irm}_j-\epsilon^{\Irm}_k}}\cdot
  \frac{(b'_1-d^{\Irm}_k)_{d^{\Irm}_j-\epsilon^{\Irm}_k}}
  {-b'_1+1+2\epsilon^{\Irm}_k}\\
  &\phantom{\phantom{(1)\ }\text{\rm W}:\ k^{[\mathcal{D},\llcN]}_a=}
  \times\prod_{\genfrac{}{}{0pt}{}{i=1}{i\neq j}}^{M_{\Irm}}
  \frac{d^{\Irm}_i-\epsilon^{\Irm}_k}{d^{\Irm}_i-d^{\Irm}_j}\,
  \frac{-b'_1+d^{\Irm}_i+\epsilon^{\Irm}_k+1}
  {-b'_1+d^{\Irm}_i+d^{\Irm}_j+1}\cdot
  \prod_{i=1}^{M_{\IIrm}}
  \frac{d^{\IIrm}_i+\epsilon^{\Irm}_k+1}{d^{\IIrm}_i+d^{\Irm}_j+1}\,
  \frac{b'_1+d^{\IIrm}_i-\epsilon^{\Irm}_k}
  {b'_1+d^{\IIrm}_i-d^{\Irm}_j},\n
  &\phantom{(1)\ }\text{\rm AW}:\ k^{[\mathcal{D},\llcN]}_a=
  \frac{h_{\mathcal{D}'_{1,jk},\llcN}}{h_{\mathcal{D},\llcN}}\,
  \frac{(a_1a_2)^{2(d^{\Irm}_j-\epsilon^{\Irm}_k)-1}
  (a_3a_4)^{-(d^{\Irm}_j-\epsilon^{\Irm}_k)}}
  {(1-q^2)(q^{\epsilon^{\Irm}_k+1};q)_{d^{\Irm}_j-\epsilon^{\Irm}_k}}\,
  \frac{q^{2-2d^{\Irm}_j(d^{\Irm}_j+1)+\epsilon^{\Irm}_k(2\epsilon^{\Irm}_k+3)}}
  {(a_1a_2q^{-d^{\Irm}_j-1},a_3a_4q^{\epsilon^{\Irm}_k};q
  )_{d^{\Irm}_j-\epsilon^{\Irm}_k}}\n
  &\phantom{\phantom{(1)\ }\text{\rm AW}:\ k^{[\mathcal{D},\llcN]}_a=}
  \times\prod_{l=1}^2\prod_{m=3}^4
  \frac{1}{(a_la_m^{-1}q^{-d^{\Irm}_j};q)_{d^{\Irm}_j-\epsilon^{\Irm}_k}}\cdot
  \frac{(b'_4q^{-d^{\Irm}_k};q)_{d^{\Irm}_j-\epsilon^{\Irm}_k}}
  {1-b^{\prime\,-1}_4q^{1+2\epsilon^{\Irm}_k}}\n
  &\phantom{\phantom{(1)\ }\text{\rm AW}:\ k^{[\mathcal{D},\llcN]}_a=}
  \times\prod_{\genfrac{}{}{0pt}{}{i=1}{i\neq j}}^{M_{\Irm}}
  \frac{1-q^{d^{\Irm}_i-\epsilon^{\Irm}_k}}{1-q^{d^{\Irm}_i-d^{\Irm}_j}}\,
  \frac{1-b^{\prime\,-1}_4q^{d^{\Irm}_i+\epsilon^{\Irm}_k+1}}
  {1-b^{\prime\,-1}_4q^{d^{\Irm}_i+d^{\Irm}_j+1}}\cdot
  \prod_{i=1}^{M_{\IIrm}}
  \frac{1-q^{d^{\IIrm}_i+\epsilon^{\Irm}_k+1}}{1-q^{d^{\IIrm}_i+d^{\Irm}_j+1}}\,
  \frac{1-b'_4q^{d^{\IIrm}_i-\epsilon^{\Irm}_k}}
  {1-b'_4q^{d^{\IIrm}_i-d^{\Irm}_j}},\n
  &(2)\ \cP_a=P_{\mathcal{D}'_{2,jk},\llcN}:\n
  &\phantom{(2)\ }\text{\rm cH}:\ k^{[\mathcal{D},\llcN]}_a=
  \frac{h_{\mathcal{D}'_{2,jk},\llcN}}{h_{\mathcal{D},\llcN}}\,
  \frac{(\epsilon^{\IIrm}_k+1)_{d^{\IIrm}_j-\epsilon^{\IIrm}_k}}{2}\,
  \frac{1}{(a_2+a_4-d^{\IIrm}_j-1,a_1+a_3+\epsilon^{\IIrm}_k
  )_{d^{\IIrm}_j-\epsilon^{\IIrm}_k}}\n
  &\phantom{\phantom{(2)\ }\text{\rm cH}:\ k^{[\mathcal{D},\llcN]}_a=}
  \times\frac{1}{(a_2-a_1-d^{\IIrm}_j)_{d^{\IIrm}_j-\epsilon^{\IIrm}_k}
  (a_4-a_3-d^{\IIrm}_j)_{d^{\IIrm}_j-\epsilon^{\IIrm}_k}}\,
  \frac{(-b^{\prime\,\text{\rm cH}}-d^{\IIrm}_k
  )_{d^{\IIrm}_j-\epsilon^{\IIrm}_k}}
  {b^{\prime\,\text{\rm cH}}+1+2\epsilon^{\IIrm}_k}\n
  &\phantom{\phantom{(2)\ }\text{\rm cH}:\ k^{[\mathcal{D},\llcN]}_a=}
  \times\prod_{\genfrac{}{}{0pt}{}{i=1}{i\neq j}}^{M_{\IIrm}}
  \frac{d^{\IIrm}_i-\epsilon^{\IIrm}_k}{d^{\IIrm}_i-d^{\IIrm}_j}\,
  \frac{b^{\prime\,\text{\rm cH}}+d^{\IIrm}_i+\epsilon^{\IIrm}_k+1}
  {b^{\prime\,\text{\rm cH}}+d^{\IIrm}_i+d^{\IIrm}_j+1}\cdot
  \prod_{i=1}^{M_{\Irm}}
  \frac{d^{\Irm}_i+\epsilon^{\IIrm}_k+1}{d^{\Irm}_i+d^{\IIrm}_j+1}\,
  \frac{-b^{\prime\,\text{\rm cH}}+d^{\Irm}_i-\epsilon^{\IIrm}_k}
  {-b^{\prime\,\text{\rm cH}}+d^{\Irm}_i-d^{\IIrm}_j},\n
  &\phantom{(2)\ }\ \text{\rm W}:\ k^{[\mathcal{D},\llcN]}_a=
  \frac{h_{\mathcal{D}'_{2,jk},\llcN}}{h_{\mathcal{D},\llcN}}\,
  \frac{1}{2(\epsilon^{\IIrm}_k+1)_{d^{\IIrm}_j-\epsilon^{\IIrm}_k}}\,
  \frac{1}{(a_3+a_4-d^{\IIrm}_j-1,a_1+a_2+\epsilon^{\IIrm}_k
  )_{d^{\IIrm}_j-\epsilon^{\IIrm}_k}}\n
  &\phantom{\phantom{(2)\ }\text{\rm W}:\ k^{[\mathcal{D},\llcN]}_a=}
  \times\prod_{l=3}^4\prod_{m=1}^2
  \frac{1}{(a_l-a_m-d^{\IIrm}_j)_{d^{\IIrm}_j-\epsilon^{\IIrm}_k}}\cdot
  \frac{(-b'_1-d^{\IIrm}_k)_{d^{\IIrm}_j-\epsilon^{\IIrm}_k}}
  {b'_1+1+2\epsilon^{\IIrm}_k}\\
  &\phantom{\phantom{(2)\ }\text{\rm W}:\ k^{[\mathcal{D},\llcN]}_a=}
  \times\prod_{\genfrac{}{}{0pt}{}{i=1}{i\neq j}}^{M_{\IIrm}}
  \frac{d^{\IIrm}_i-\epsilon^{\IIrm}_k}{d^{\IIrm}_i-d^{\IIrm}_j}\,
  \frac{b'_1+d^{\IIrm}_i+\epsilon^{\IIrm}_k+1}
  {b'_1+d^{\IIrm}_i+d^{\IIrm}_j+1}\cdot
  \prod_{i=1}^{M_{\Irm}}
  \frac{d^{\Irm}_i+\epsilon^{\IIrm}_k+1}{d^{\Irm}_i+d^{\IIrm}_j+1}\,
  \frac{-b'_1+d^{\Irm}_i-\epsilon^{\IIrm}_k}
  {-b'_1+d^{\Irm}_i-d^{\IIrm}_j},\n
  &\phantom{(2)\ }\text{\rm AW}:\ k^{[\mathcal{D},\llcN]}_a=
  \frac{h_{\mathcal{D}'_{2,jk},\llcN}}{h_{\mathcal{D},\llcN}}\,
  \frac{(a_3a_4)^{2(d^{\IIrm}_j-\epsilon^{\IIrm}_k)-1}
  (a_1a_2)^{-(d^{\IIrm}_j-\epsilon^{\IIrm}_k)}}
  {(1-q^2)(q^{\epsilon^{\IIrm}_k+1};q)_{d^{\IIrm}_j-\epsilon^{\IIrm}_k}}\,
  \frac{q^{2-2d^{\IIrm}_j(d^{\IIrm}_j+1)
  +\epsilon^{\IIrm}_k(2\epsilon^{\IIrm}_k+3)}}
  {(a_3a_4q^{-d^{\IIrm}_j-1},a_1a_2q^{\epsilon^{\IIrm}_k};q
  )_{d^{\IIrm}_j-\epsilon^{\IIrm}_k}}\n
  &\phantom{\phantom{(2)\ }\text{\rm AW}:\ k^{[\mathcal{D},\llcN]}_a=}
  \times\prod_{l=3}^4\prod_{m=1}^2
  \frac{1}{(a_la_m^{-1}q^{-d^{\IIrm}_j};q)_{d^{\IIrm}_j-\epsilon^{\IIrm}_k}}\cdot
  \frac{(b^{\prime\,-1}_4q^{-d^{\IIrm}_k};q)_{d^{\IIrm}_j-\epsilon^{\IIrm}_k}}
  {1-b'_4q^{1+2\epsilon^{\IIrm}_k}}\n
  &\phantom{\phantom{(2)\ }\text{\rm AW}:\ k^{[\mathcal{D},\llcN]}_a=}
  \times\prod_{\genfrac{}{}{0pt}{}{i=1}{i\neq j}}^{M_{\IIrm}}
  \frac{1-q^{d^{\IIrm}_i-\epsilon^{\IIrm}_k}}{1-q^{d^{\IIrm}_i-d^{\IIrm}_j}}\,
  \frac{1-b'_4q^{d^{\IIrm}_i+\epsilon^{\IIrm}_k+1}}
  {1-b'_4q^{d^{\IIrm}_i+d^{\IIrm}_j+1}}\cdot
  \prod_{i=1}^{M_{\Irm}}
  \frac{1-q^{d^{\Irm}_i+\epsilon^{\IIrm}_k+1}}{1-q^{d^{\Irm}_i+d^{\IIrm}_j+1}}\,
  \frac{1-b^{\prime\,-1}_4q^{d^{\Irm}_i-\epsilon^{\IIrm}_k}}
  {1-b^{\prime\,-1}_4q^{d^{\Irm}_i-d^{\IIrm}_j}},\n
  &(3)\ \cP_a=P_{\mathcal{D}'_{3,jk},\llcN}:\n
  &\phantom{(3)\ }\text{\rm cH}:\ k^{[\mathcal{D},\llcN]}_a=
  \frac{h_{\mathcal{D}'_{3,jk},\llcN}}{h_{\mathcal{D},\llcN}}\,
  \frac{(-1)^{d^{\Irm}_j+d^{\IIrm}_k+1}d^{\Irm}_j!\,d^{\IIrm}_k!}
  {2(d^{\Irm}_j+d^{\IIrm}_k+1)}\,
  \frac{1}{(a_1+a_3-d^{\Irm}_j-1,a_2+a_4-d^{\IIrm}_k-1
  )_{d^{\Irm}_j+d^{\IIrm}_k+1}}\n
  &\phantom{\phantom{(3)\ }\text{\rm cH}:\ k^{[\mathcal{D},\llcN]}_a=}
  \times\frac{(-b^{\prime\,\text{\rm cH}}-d^{\IIrm}_k)_{d^{\Irm}_j}
  (b^{\prime\,\text{\rm cH}}-d^{\Irm}_j)_{d^{\IIrm}_k}}
  {(a_1-a_2-d^{\Irm}_j,a_3-a_4-d^{\Irm}_j)_{d^{\Irm}_j+d^{\IIrm}_k+1}}\n
  &\phantom{\phantom{(3)\ }\text{\rm cH}:\ k^{[\mathcal{D},\llcN]}_a=}
  \times\prod_{\genfrac{}{}{0pt}{}{i=1}{i\neq j}}^{M_{\Irm}}
  \frac{1}{(d^{\Irm}_i-d^{\Irm}_j)(d^{\Irm}_i+d^{\IIrm}_k+1)
  (-b^{\prime\,\text{\rm cH}}+d^{\Irm}_i+d^{\Irm}_j+1)
  (-b^{\prime\,\text{\rm cH}}+d^{\Irm}_i-d^{\IIrm}_k)}\n
  &\phantom{\phantom{(3)\ }\text{\rm cH}:\ k^{[\mathcal{D},\llcN]}_a=}
  \times\prod_{\genfrac{}{}{0pt}{}{i=1}{i\neq k}}^{M_{\IIrm}}
  \frac{1}{(d^{\IIrm}_i-d^{\IIrm}_k)(d^{\IIrm}_i+d^{\Irm}_j+1)
  (b^{\prime\,\text{\rm cH}}+d^{\IIrm}_i+d^{\IIrm}_k+1)
  (b^{\prime\,\text{\rm cH}}+d^{\IIrm}_i-d^{\Irm}_j)},\n
  &\phantom{(3)\ }\ \text{\rm W}:\ k^{[\mathcal{D},\llcN]}_a=
  \frac{h_{\mathcal{D}'_{3,jk},\llcN}}{h_{\mathcal{D},\llcN}}\,
  \frac{(-1)^{d^{\Irm}_j+d^{\IIrm}_k+1}}
  {2(d^{\Irm}_j+d^{\IIrm}_k+1)d^{\Irm}_j!\,d^{\IIrm}_k!}\,
  \frac{1}{(a_1+a_2-d^{\Irm}_j-1,a_3+a_4-d^{\IIrm}_k-1
  )_{d^{\Irm}_j+d^{\IIrm}_k+1}}\n
  &\phantom{\phantom{(3)\ }\text{\rm W}:\ k^{[\mathcal{D},\llcN]}_a=}
  \times\frac{(-b'_1-d^{\IIrm}_k)_{d^{\Irm}_j}(b'_1-d^{\Irm}_j)_{d^{\IIrm}_k}}
  {\prod_{l=1}^2\prod_{m=3}^4(a_l-a_m-d^{\Irm}_j)_{d^{\Irm}_j+d^{\IIrm}_k+1}}\n
  &\phantom{\phantom{(3)\ }\text{\rm W}:\ k^{[\mathcal{D},\llcN]}_a=}
  \times\prod_{\genfrac{}{}{0pt}{}{i=1}{i\neq j}}^{M_{\Irm}}
  \frac{1}{(d^{\Irm}_i-d^{\Irm}_j)(d^{\Irm}_i+d^{\IIrm}_k+1)
  (-b'_1+d^{\Irm}_i+d^{\Irm}_j+1)(-b'_1+d^{\Irm}_i-d^{\IIrm}_k)}\\
  &\phantom{\phantom{(3)\ }\text{\rm W}:\ k^{[\mathcal{D},\llcN]}_a=}
  \times\prod_{\genfrac{}{}{0pt}{}{i=1}{i\neq k}}^{M_{\IIrm}}
  \frac{1}{(d^{\IIrm}_i-d^{\IIrm}_k)(d^{\IIrm}_i+d^{\Irm}_j+1)
  (b'_1+d^{\IIrm}_i+d^{\IIrm}_k+1)(b'_1+d^{\IIrm}_i-d^{\Irm}_j)},\n
  &\phantom{(3)\ }\text{\rm AW}:\ k^{[\mathcal{D},\llcN]}_a=
  \frac{h_{\mathcal{D}'_{3,jk},\llcN}}{h_{\mathcal{D},\llcN}}\,
  \frac{(-1)^{d^{\Irm}_j+d^{\IIrm}_k+1}(a_1a_2)^{3d^{\Irm}_j+2}
  (a_3a_4)^{d^{\IIrm}_k-2d^{\Irm}_j}}
  {(1-q^2)(1-q^{d^{\Irm}_j+d^{\IIrm}_k+1})
  (q;q)_{d^{\Irm}_j}(q;q)_{d^{\IIrm}_k}}\n
  &\phantom{\phantom{(3)\ }\text{\rm AW}:\ k^{[\mathcal{D},\llcN]}_a=}
  \times\frac{q^{-\frac12(5d^{\Irm\,2}_j-2d^{\Irm}_jd^{\IIrm}_k+d^{\IIrm\,2}_k
  +3d^{\Irm}_j-d^{\IIrm}_k+8)}}
  {(a_1a_2q^{-d^{\Irm}_j-1},a_3a_4q^{-d^{\IIrm}_k-1};q
  )_{d^{\Irm}_j+d^{\IIrm}_k+1}}\,
  \frac{(b^{\prime\,-1}_4q^{-d^{\IIrm}_k};q)_{d^{\Irm}_j}
  (b'_4q^{-d^{\Irm}_j};q)_{d^{\IIrm}_k}}
  {\prod_{l=1}^2\prod_{m=3}^4(a_la_m^{-1}q^{-d^{\Irm}_j};q
  )_{d^{\Irm}_j+d^{\IIrm}_k+1}}\n
  &\phantom{\phantom{(3)\ }\text{\rm AW}:\ k^{[\mathcal{D},\llcN]}_a=}
  \times\prod_{\genfrac{}{}{0pt}{}{i=1}{i\neq j}}^{M_{\Irm}}
  \frac{b^{\prime\,-1}_4q^{2(d^{\Irm}_i-i-M_{\IIrm})-1}}
  {(1-q^{d^{\Irm}_i-d^{\Irm}_j})(1-q^{d^{\Irm}_i+d^{\IIrm}_k+1})
  (1-b^{\prime\,-1}_4q^{d^{\Irm}_i+d^{\Irm}_j+1})
  (1-b^{\prime\,-1}_4q^{d^{\Irm}_i-d^{\IIrm}_k})}\n
  &\phantom{\phantom{(3)\ }\text{\rm AW}:\ k^{[\mathcal{D},\llcN]}_a=}
  \times\prod_{\genfrac{}{}{0pt}{}{i=1}{i\neq k}}^{M_{\IIrm}}
  \frac{b'_4q^{2(d^{\IIrm}_i-i-M_{\Irm})-1}}
  {(1-q^{d^{\IIrm}_i-d^{\IIrm}_k})(1-q^{d^{\IIrm}_i+d^{\Irm}_j+1})
  (1-b'_4q^{d^{\IIrm}_i+d^{\IIrm}_k+1})
  (1-b'_4q^{d^{\IIrm}_i-d^{\Irm}_j})},\nonumber
\end{align}
where $b^{\prime\,\text{\rm cH}}_1=a_1+a_3-a_2-a_4$, $b'_1=a_1+a_2-a_3-a_4$ and
$b'_4=a_1a_2a_3^{-1}a_4^{-1}$.
\end{conj}
By numerical calculation, we can verify this conjecture for small $M$, $d_j$,
$\cN$ and $n$.

\section{Summary and Comments}
\label{sec:summary}

Ordinary orthogonal polynomials satisfy the discrete orthogonal relations
\eqref{doorg}.
For the multi-indexed orthogonal polynomials in oQM, which satisfy the second
order differential equations, the discrete orthogonal relations also hold
\cite{hs19}.
We have generalized this result to the multi-indexed orthogonal polynomials in
idQM, which satisfy the second order difference equations.
The discrete orthogonal relations hold for the case-(1) multi-indexed orthogonal
polynomials of continuous Hahn, Wilson and Askey-Wilson types,
Theorem\,\ref{thm:domiop}, and their normalization constants are conjectured,
Conjecture\,\ref{conj:kDNa}.
We remark that, as in \cite{hs19}, the discrete orthogonal relations can also
be shown for the Krein-Adler type multi-indexed orthogonal polynomials
in idQM. Since its derivation is similar to \S\,\ref{sec:miop} and \cite{hs19},
we omit it.
Its detailed calculations can be found in arXiv:2112.09358\underline{v1},
where the conjectures for the normalization constants of the discrete
orthogonal relations studied in \cite{hs19} are also given.
We hope that the conjectures for the normalization constants will be proved.

In addition to oQM and idQM, we have another quantum mechanical system, rdQM.
Orthogonal polynomials appearing in rdQM are the $q$-Racah polynomial, its
various limits, and their multi-indexed versions. The discrete orthogonal
relations are expected to hold for the multi-indexed orthogonal polynomials in
rdQM as well.
It is an interesting problem to study these discrete orthogonal relations
concretely.

\section*{Acknowledgements}

I thank Ryu Sasaki for discussions in the early stage of this work.
This work is supported by JSPS KAKENHI Grant Number JP19K03667.

\bigskip
\appendix
\section{Proof of Proposition\,\ref{prop:id}}
\label{app:pr:id}

In this appendix we prove Proposition\,\ref{prop:id}.
We use the following abbreviated notations:
\begin{alignat}{2}
  w(x)&=\text{W}_{\gamma}[\tilde{\phi}_{d_1},\ldots,\tilde{\phi}_{d_M}](x),
  &\quad u(x)&=\text{W}_{\gamma}[\tilde{\phi}_{d_1},\ldots,\tilde{\phi}_{d_M},
  \phi_n](x),\n
  w_1(x)&=\text{W}_{\gamma}[\tilde{\phi}_{d_1},\ldots,\tilde{\phi}_{d_M},
  \tilde{\phi}_{d'}](x),
  &\quad u_1(x)&=\text{W}_{\gamma}[\tilde{\phi}_{d_1},\ldots,\tilde{\phi}_{d_M},
  \tilde{\phi}_{d'},\phi_n](x),\n
  w_2(x)&=\text{W}_{\gamma}[\tilde{\phi}_{d_1},\ldots,\tilde{\phi}_{d_M},
  \tilde{\phi}_{d''}](x),
  &\quad u_2(x)&=\text{W}_{\gamma}[\tilde{\phi}_{d_1},\ldots,\tilde{\phi}_{d_M},
  \tilde{\phi}_{d''},\phi_n](x),
  \label{wu}\\
  w_3(x)&=\text{W}_{\gamma}[\tilde{\phi}_{d_1},\ldots,\tilde{\phi}_{d_M},
  \tilde{\phi}_{d'},\tilde{\phi}_{d''}](x),
  &\quad u_3(x)&=\text{W}_{\gamma}[\tilde{\phi}_{d_1},\ldots,\tilde{\phi}_{d_M},
  \tilde{\phi}_{d'},\tilde{\phi}_{d''},\phi_n](x).
  \nonumber
\end{alignat}
For example, we have
\begin{align*}
  \hat{V}_{\mathcal{D}''}(x)
  &=\sqrt{V(x-i\tfrac{M}{2}\gamma)V^*(x-i\tfrac{M+2}{2}\gamma)}\,
  \frac{w(x+i\frac{\gamma}{2})}{w(x-i\frac{\gamma}{2})}
  \frac{w_2(x-i\gamma)}{w_2(x)},\\
  \phi_{\mathcal{D}''\,n}(x)
  &=\Bigl(\prod_{j=0}^{M}V\bigl(x+i(\tfrac{M+1}{2}-j)\gamma\bigr)
  V^*\bigl(x-i(\tfrac{M+1}{2}-j)\gamma\bigr)\Bigr)^{\frac14}
  \frac{u_2(x)}{\sqrt{w_2(x-i\tfrac{\gamma}{2})w_2(x+i\tfrac{\gamma}{2})}}\n
  &=c^{\phi}_{\mathcal{D}''}\psi_{\mathcal{D}''}(x)
  \check{P}_{\mathcal{D}'',n}(x)
  =c^{\phi}_{\mathcal{D}''}\psi_{\mathcal{D}''}(x)
  g^P_{\mathcal{D}''}(x)^{-1}u_2(x).
\end{align*}

\subsection{Proof of Proposition\,\ref{prop:id} (1)}
\label{app:pr:id(1)}

{}From $\mathcal{H}_{\mathcal{D}''}=\hat{\mathcal{A}}_{\mathcal{D}''}
\hat{\mathcal{A}}_{\mathcal{D}''}^{\dagger}+\tilde{\mathcal{E}}_{d''}$ and
\begin{align}
  \widetilde{\mathcal{H}}'_{\mathcal{D}''}
  &\eqdef g^P_{\mathcal{D}''}(x)\circ\widetilde{\mathcal{H}}_{\mathcal{D}''}
  \circ g^P_{\mathcal{D}''}(x)^{-1}
  =g^P_{\mathcal{D}''}(x)\psi_{\mathcal{D}''}(x)^{-1}
  \circ\mathcal{H}_{\mathcal{D}''}\circ
  \psi_{\mathcal{D}''}(x)g^P_{\mathcal{D}''}(x)^{-1}\n
  &=\Bigl(\prod_{j=0}^MV\bigl(x+i(\tfrac{M+1}{2}-j)\gamma\bigr)
  V^*\bigl(x-i(\tfrac{M+1}{2}-j)\gamma\bigr)\Bigr)^{-\frac14}
  \sqrt{w_2(x-i\tfrac{\gamma}{2})w_2(x+i\tfrac{\gamma}{2})}\n
  &\quad\circ\mathcal{H}_{\mathcal{D}''}\circ
  \Bigl(\prod_{j=0}^MV\bigl(x+i(\tfrac{M+1}{2}-j)\gamma\bigr)
  V^*\bigl(x-i(\tfrac{M+1}{2}-j)\gamma\bigr)\Bigr)^{\frac14}
  \frac{1}{\sqrt{w_2(x-i\tfrac{\gamma}{2})w_2(x+i\tfrac{\gamma}{2})}}\n
  &=\sqrt{V(x-i\tfrac{M+1}{2}\gamma)V^*(x-i\tfrac{M+3}{2}\gamma)}\,
  \frac{w_2(x+i\frac{\gamma}{2})}{w_2(x-i\frac{\gamma}{2})}\,e^{\gamma p}\n
  &\quad+\sqrt{V(x+i\tfrac{M+3}{2}\gamma)V^*(x+i\tfrac{M+1}{2}\gamma)}\,
  \frac{w_2(x-i\frac{\gamma}{2})}{w_2(x+i\frac{\gamma}{2})}\,e^{-\gamma p}\n
  &\quad-\sqrt{V(x-i\tfrac{M-1}{2}\gamma)V^*(x-i\tfrac{M+1}{2}\gamma)}\,
  \frac{w(x+i\gamma)}{w(x)}
  \frac{w_2(x-i\frac{\gamma}{2})}{w_2(x+i\frac{\gamma}{2})}\n
  &\quad-\sqrt{V(x+i\tfrac{M+1}{2}\gamma)V^*(x+i\tfrac{M-1}{2}\gamma)}\,
  \frac{w(x-i\gamma)}{w(x)}
  \frac{w_2(x+i\frac{\gamma}{2})}{w_2(x-i\frac{\gamma}{2})}
  +\tilde{\mathcal{E}}_{d''},
  \label{tH'D''}
\end{align}
we have
\begin{align}
  \widetilde{\mathcal{H}}'_{\mathcal{D}''}u_1(x)
  &=\sqrt{V(x-i\tfrac{M+1}{2}\gamma)V^*(x-i\tfrac{M+3}{2}\gamma)}\,
  \frac{w_2(x+i\frac{\gamma}{2})}{w_2(x-i\frac{\gamma}{2})}\,u_1(x-i\gamma)\n
  &\quad+\sqrt{V(x+i\tfrac{M+3}{2}\gamma)V^*(x+i\tfrac{M+1}{2}\gamma)}\,
  \frac{w_2(x-i\frac{\gamma}{2})}{w_2(x+i\frac{\gamma}{2})}\,u_1(x+i\gamma)\n
  &\quad-\sqrt{V(x-i\tfrac{M-1}{2}\gamma)V^*(x-i\tfrac{M+1}{2}\gamma)}\,
  \frac{w(x+i\gamma)}{w(x)}
  \frac{w_2(x-i\frac{\gamma}{2})}{w_2(x+i\frac{\gamma}{2})}\,u_1(x)
  \label{id_i}\\
  &\quad-\sqrt{V(x+i\tfrac{M+1}{2}\gamma)V^*(x+i\tfrac{M-1}{2}\gamma)}\,
  \frac{w(x-i\gamma)}{w(x)}
  \frac{w_2(x+i\frac{\gamma}{2})}{w_2(x-i\frac{\gamma}{2})}\,u_1(x)
  +\tilde{\mathcal{E}}_{d''}\,u_1(x).
  \nonumber
\end{align}
{}From $\mathcal{H}_{\mathcal{D}'}=\hat{\mathcal{A}}_{\mathcal{D}'}
\hat{\mathcal{A}}_{\mathcal{D}'}^{\dagger}+\tilde{\mathcal{E}}_{d'}$ and
$\mathcal{H}_{\mathcal{D}'}\phi_{\mathcal{D}'\,n}(x)
=\mathcal{E}_n\phi_{\mathcal{D}'\,n}(x)$, we have
\begin{align}
  \mathcal{E}_nu_1(x)
  &=\sqrt{V(x-i\tfrac{M+1}{2}\gamma)V^*(x-i\tfrac{M+3}{2}\gamma)}\,
  \frac{w_1(x+i\frac{\gamma}{2})}{w_1(x-i\frac{\gamma}{2})}\,u_1(x-i\gamma)\n
  &\quad+\sqrt{V(x+i\tfrac{M+3}{2}\gamma)V^*(x+i\tfrac{M+1}{2}\gamma)}\,
  \frac{w_1(x-i\frac{\gamma}{2})}{w_1(x+i\frac{\gamma}{2})}\,u_1(x+i\gamma)\n
  &\quad-\sqrt{V(x-i\tfrac{M-1}{2}\gamma)V^*(x-i\tfrac{M+1}{2}\gamma)}\,
  \frac{w(x+i\gamma)}{w(x)}
  \frac{w_1(x-i\frac{\gamma}{2})}{w_1(x+i\frac{\gamma}{2})}\,u_1(x)\n
  &\quad-\sqrt{V(x+i\tfrac{M+1}{2}\gamma)V^*(x+i\tfrac{M-1}{2}\gamma)}\,
  \frac{w(x-i\gamma)}{w(x)}
  \frac{w_1(x+i\frac{\gamma}{2})}{w_1(x-i\frac{\gamma}{2})}\,u_1(x)
  +\tilde{\mathcal{E}}_{d'}\,u_1(x).
  \label{id_a}
\end{align}
By rewriting these \eqref{id_a} and \eqref{id_i} as
\begin{equation}
  \begin{pmatrix}
  \alpha_1a_1&\alpha_1^*a_1^*\\
  \alpha_1a_2&\alpha_1^*a_2^*
  \end{pmatrix}
  \begin{pmatrix}
  u_1(x-i\gamma)\\
  u_1(x+i\gamma)
  \end{pmatrix}
  =\begin{pmatrix}
  b_1u_1(x)\\
  b_2u_1(x)
 \end{pmatrix},
\end{equation}
where
\begin{align*}
  &\alpha_1=\sqrt{V(x-i\tfrac{M+1}{2}\gamma)V^*(x-i\tfrac{M+3}{2}\gamma)},\quad
  \alpha_2=\sqrt{V(x-i\tfrac{M-1}{2}\gamma)V^*(x-i\tfrac{M+1}{2}\gamma)},\\
  &a_1=\frac{w_1(x+i\frac{\gamma}{2})}{w_1(x-i\frac{\gamma}{2})}
  =\frac{1}{a_1^*},\quad
  a_2=\frac{w_2(x+i\frac{\gamma}{2})}{w_2(x-i\frac{\gamma}{2})}
  =\frac{1}{a_2^*},\quad
  c=\frac{w(x+i\gamma)}{w(x)},\\
  &b_1=\mathcal{E}_n-\tilde{\mathcal{E}}_{d'}
  +\alpha_2ca_1^*+\alpha_2^*c^*a_1,\quad
  b_2=\widetilde{\mathcal{H}}'_{\mathcal{D}''}
  -\tilde{\mathcal{E}}_{d''}
  +\alpha_2ca_2^*+\alpha_2^*c^*a_2,
\end{align*}
we obtain
\begin{equation}
  \begin{pmatrix}
  \alpha_1u_1(x-i\gamma)\\
  \alpha_1^*u_1(x+i\gamma)
  \end{pmatrix}
  =\frac{1}{a_1a_2^*-a_1^*a_2}
  \begin{pmatrix}
  a_2^*&-a_1^*\\
  -a_2&a_1
  \end{pmatrix}
  \begin{pmatrix}
  b_1u_1(x)\\
  b_2u_1(x)
 \end{pmatrix}.
  \label{u1byu1}
\end{equation}

By writing down the identity \eqref{idmaru1} and multiplying by
\begin{equation*}
  (\mathcal{E}_n-\tilde{\mathcal{E}}_{d'})
  \Bigl(\prod_{j=0}^MV\bigl(x+i(\tfrac{M+1}{2}-j)\gamma\bigr)
  V^*\bigl(x-i(\tfrac{M+1}{2}-j)\gamma\bigr)\Bigr)^{-\frac14}
  \sqrt{w_2(x-i\tfrac{\gamma}{2})w_2(x+i\tfrac{\gamma}{2})},
\end{equation*}
we obtain
\begin{align}
  (\mathcal{E}_n-\tilde{\mathcal{E}}_{d'})u_2(x)
  &=\alpha_1\frac{w_2(x+i\frac{\gamma}{2})}{w_1(x-i\frac{\gamma}{2})}
  u_1(x-i\gamma)
  +\alpha_1^*\frac{w_2(x-i\frac{\gamma}{2})}{w_1(x+i\frac{\gamma}{2})}
  u_1(x+i\gamma)\n
  &\quad-\alpha_2\frac{w_2(x-i\frac{\gamma}{2})}{w_1(x+i\frac{\gamma}{2})}
  cu_1(x)
  -\alpha_2^*\frac{w_2(x+i\frac{\gamma}{2})}{w_1(x-i\frac{\gamma}{2})}
  c^*u_1(x).
  \label{idmaru1u}
\end{align}
By substituting $u_1(x\mp i\gamma)$ in \eqref{u1byu1} into \eqref{idmaru1u},
we obtain
\begin{equation}
  (a_1a_2^*-a_1^*a_2)(\mathcal{E}_n-\tilde{\mathcal{E}}_{d'})u_2(x)
  =\Bigl(\frac{w_2(x-i\frac{\gamma}{2})}{w_1(x-i\frac{\gamma}{2})}
  -\frac{w_2(x+i\frac{\gamma}{2})}{w_1(x+i\frac{\gamma}{2})}\Bigr)
  \bigl(\widetilde{\mathcal{H}}'_{\mathcal{D}''}u_1(x)
  +(\mathcal{E}_n-\tilde{\mathcal{E}}_{d'}-\tilde{\mathcal{E}}_{d''})u_1(x)
  \bigr).
  \label{idmaru1u2}
\end{equation}
Since $u_1(x)=g^P_{\mathcal{D}'}(x)\check{P}_{\mathcal{D}',n}(x)$, we have
\begin{align*}
  &\quad\widetilde{\mathcal{H}}'_{\mathcal{D}''}u_1(x)
  +(\mathcal{E}_n-\tilde{\mathcal{E}}_{d'}-\tilde{\mathcal{E}}_{d''})u_1(x)\\
  &=g^P_{\mathcal{D}''}(x)\bigl(g^P_{\mathcal{D}''}(x)^{-1}\circ
  \widetilde{\mathcal{H}}'_{\mathcal{D}''}\circ g^P_{\mathcal{D}''}(x)
  +\mathcal{E}_n-\tilde{\mathcal{E}}_{d'}-\tilde{\mathcal{E}}_{d''}\bigr)
  \frac{g^P_{\mathcal{D}'}(x)}{g^P_{\mathcal{D}''}(x)}
  \check{P}_{\mathcal{D}',n}(x)\\
  &=g^P_{\mathcal{D}''}(x)(\widetilde{\mathcal{H}}_{\mathcal{D}''}
  +\mathcal{E}_n-\tilde{\mathcal{E}}_{d'}-\tilde{\mathcal{E}}_{d''})
  \frac{g^P_{\mathcal{D}'}(x)}{g^P_{\mathcal{D}''}(x)}
  \check{P}_{\mathcal{D}',n}(x).
\end{align*}
Therefore \eqref{idmaru1u2} gives
\begin{equation*}
  \Bigl(\frac{w_1(x-i\frac{\gamma}{2})}{w_2(x-i\frac{\gamma}{2})}
  +\frac{w_1(x+i\frac{\gamma}{2})}{w_2(x+i\frac{\gamma}{2})}\Bigr)
  (\mathcal{E}_n-\tilde{\mathcal{E}}_{d'})\check{P}_{\mathcal{D}'',n}(x)
  =(\widetilde{\mathcal{H}}_{\mathcal{D}''}
  +\mathcal{E}_n-\tilde{\mathcal{E}}_{d'}-\tilde{\mathcal{E}}_{d''})
  \frac{g^P_{\mathcal{D}'}(x)}{g^P_{\mathcal{D}''}(x)}
  \check{P}_{\mathcal{D}',n}(x),
\end{equation*}
namely,
\begin{align}
  &\quad(\mathcal{E}_n-\tilde{\mathcal{E}}_{d'})
  \Bigl(\frac{g_{\mathcal{D}'}(x-i\frac{\gamma}{2})}
  {g_{\mathcal{D}''}(x-i\frac{\gamma}{2})}
  \frac{\check{\Xi}_{\mathcal{D}'}(x-i\frac{\gamma}{2})}
  {\check{\Xi}_{\mathcal{D}''}(x-i\frac{\gamma}{2})}
  +\frac{g_{\mathcal{D}'}(x+i\frac{\gamma}{2})}
  {g_{\mathcal{D}''}(x+i\frac{\gamma}{2})}
  \frac{\check{\Xi}_{\mathcal{D}'}(x+i\frac{\gamma}{2})}
  {\check{\Xi}_{\mathcal{D}''}(x+i\frac{\gamma}{2})}\Bigr)
  \check{P}_{\mathcal{D}'',n}(x)\n
  &=(\widetilde{\mathcal{H}}_{\mathcal{D}''}
  +\mathcal{E}_n-\tilde{\mathcal{E}}_{d'}-\tilde{\mathcal{E}}_{d''})
  \frac{g^P_{\mathcal{D}'}(x)}{g^P_{\mathcal{D}''}(x)}
  \check{P}_{\mathcal{D}',n}(x).
  \label{idmaru1P2}
\end{align}
When the types of $d'$ and $d''$ are the same, we have
$g_{\mathcal{D}'}(x)=g_{\mathcal{D}''}(x)$ and
$g^P_{\mathcal{D}'}(x)=g^P_{\mathcal{D}''}(x)$.
Thus \eqref{idmaru1P2} gives \eqref{idmaru1P}.
\hfill\fbox{}

\subsection{Proof of Proposition\,\ref{prop:id} (2)}
\label{app:pr:id(2)}

Like as \eqref{tH'D''}, $\widetilde{\mathcal{H}}'_{\mathcal{D}'''}
\eqdef g^P_{\mathcal{D}'''}(x)\circ\widetilde{\mathcal{H}}_{\mathcal{D}'''}
\circ g^P_{\mathcal{D}'''}(x)^{-1}$ is expressed as
\begin{align*}
  \widetilde{\mathcal{H}}'_{\mathcal{D}'''}
  &=\sqrt{V(x-i\tfrac{M+2}{2}\gamma)V^*(x-i\tfrac{M+4}{2}\gamma)}\,
  \frac{w_3(x+i\frac{\gamma}{2})}{w_3(x-i\frac{\gamma}{2})}\,e^{\gamma p}\\
  &\quad+\sqrt{V(x+i\tfrac{M+4}{2}\gamma)V^*(x+i\tfrac{M+2}{2}\gamma)}\,
  \frac{w_3(x-i\frac{\gamma}{2})}{w_3(x+i\frac{\gamma}{2})}\,e^{-\gamma p}\\
  &\quad-\sqrt{V(x-i\tfrac{M}{2}\gamma)V^*(x-i\tfrac{M+2}{2}\gamma)}\,
  \frac{w_1(x+i\gamma)}{w_1(x)}
  \frac{w_3(x-i\frac{\gamma}{2})}{w_3(x+i\frac{\gamma}{2})}\\
  &\quad-\sqrt{V(x+i\tfrac{M+2}{2}\gamma)V^*(x+i\tfrac{M}{2}\gamma)}\,
  \frac{w_1(x-i\gamma)}{w_1(x)}
  \frac{w_3(x+i\frac{\gamma}{2})}{w_3(x-i\frac{\gamma}{2})}
  +\tilde{\mathcal{E}}_{d''}.
\end{align*}
Then we have
\begin{align*}
  &\quad g^P_{\mathcal{D}}(x)
  \circ\widetilde{\mathcal{H}}_{\mathcal{D}'''}\circ g^P_{\mathcal{D}}(x)^{-1}
  =\frac{g^P_{\mathcal{D}}(x)}{g^P_{\mathcal{D}'''}(x)}
  \circ\widetilde{\mathcal{H}}'_{\mathcal{D}'''}\circ
  \frac{g^P_{\mathcal{D}'''}(x)}{g^P_{\mathcal{D}}(x)}\\
  &=\sqrt{V(x-i\tfrac{M+2}{2}\gamma)V^*(x-i\tfrac{M+4}{2}\gamma)}\,
  \frac{w_3(x+i\frac{\gamma}{2})}{w_3(x-i\frac{\gamma}{2})}
  \frac{g^P_{\mathcal{D}}(x)}{g^P_{\mathcal{D}}(x-i\gamma)}
  \frac{g^P_{\mathcal{D}'''}(x-i\gamma)}{g^P_{\mathcal{D}'''}(x)}
  \,e^{\gamma p}\\
  &\quad+\sqrt{V(x+i\tfrac{M+4}{2}\gamma)V^*(x+i\tfrac{M+2}{2}\gamma)}\,
  \frac{w_3(x-i\frac{\gamma}{2})}{w_3(x+i\frac{\gamma}{2})}
  \frac{g^P_{\mathcal{D}}(x)}{g^P_{\mathcal{D}}(x+i\gamma)}
  \frac{g^P_{\mathcal{D}'''}(x+i\gamma)}{g^P_{\mathcal{D}'''}(x)}
  \,e^{-\gamma p}\\
  &\quad-\sqrt{V(x-i\tfrac{M}{2}\gamma)V^*(x-i\tfrac{M+2}{2}\gamma)}\,
  \frac{w_1(x+i\gamma)}{w_1(x)}
  \frac{w_3(x-i\frac{\gamma}{2})}{w_3(x+i\frac{\gamma}{2})}\\
  &\quad-\sqrt{V(x+i\tfrac{M+2}{2}\gamma)V^*(x+i\tfrac{M}{2}\gamma)}\,
  \frac{w_1(x-i\gamma)}{w_1(x)}
  \frac{w_3(x+i\frac{\gamma}{2})}{w_3(x-i\frac{\gamma}{2})}
  +\tilde{\mathcal{E}}_{d''},
\end{align*}
and obtain
\begin{align}
  &\quad\bigl(g^P_{\mathcal{D}}(x)
  \circ\widetilde{\mathcal{H}}_{\mathcal{D}'''}\circ g^P_{\mathcal{D}}(x)^{-1}
  \bigr)u(x)\n
  &=\sqrt{V(x-i\tfrac{M+2}{2}\gamma)V^*(x-i\tfrac{M+4}{2}\gamma)}\,
  \frac{w_3(x+i\frac{\gamma}{2})}{w_3(x-i\frac{\gamma}{2})}
  \frac{g^P_{\mathcal{D}}(x)}{g^P_{\mathcal{D}}(x-i\gamma)}
  \frac{g^P_{\mathcal{D}'''}(x-i\gamma)}{g^P_{\mathcal{D}'''}(x)}
  \,u(x-i\gamma)\n
  &\quad+\sqrt{V(x+i\tfrac{M+4}{2}\gamma)V^*(x+i\tfrac{M+2}{2}\gamma)}\,
  \frac{w_3(x-i\frac{\gamma}{2})}{w_3(x+i\frac{\gamma}{2})}
  \frac{g^P_{\mathcal{D}}(x)}{g^P_{\mathcal{D}}(x+i\gamma)}
  \frac{g^P_{\mathcal{D}'''}(x+i\gamma)}{g^P_{\mathcal{D}'''}(x)}
  \,u(x+i\gamma)\n
  &\quad-\sqrt{V(x-i\tfrac{M}{2}\gamma)V^*(x-i\tfrac{M+2}{2}\gamma)}\,
  \frac{w_1(x+i\gamma)}{w_1(x)}
  \frac{w_3(x-i\frac{\gamma}{2})}{w_3(x+i\frac{\gamma}{2})}\,u(x)\n
  &\quad-\sqrt{V(x+i\tfrac{M+2}{2}\gamma)V^*(x+i\tfrac{M}{2}\gamma)}\,
  \frac{w_1(x-i\gamma)}{w_1(x)}
  \frac{w_3(x+i\frac{\gamma}{2})}{w_3(x-i\frac{\gamma}{2})}\,u(x)
  +\tilde{\mathcal{E}}_{d''}\,u(x).
  \label{id_i2}
\end{align}
{}From $\mathcal{H}_{\mathcal{D}}=\hat{\mathcal{A}}_{\mathcal{D}'}^{\dagger}
\hat{\mathcal{A}}_{\mathcal{D}'}+\tilde{\mathcal{E}}_{d'}$ and
$\mathcal{H}_{\mathcal{D}}\phi_{\mathcal{D}\,n}(x)
=\mathcal{E}_n\phi_{\mathcal{D}\,n}(x)$, we have
\begin{align}
  \mathcal{E}_nu(x)
  &=\sqrt{V(x-i\tfrac{M}{2}\gamma)V^*(x-i\tfrac{M+2}{2}\gamma)}\,
  \frac{w(x+i\frac{\gamma}{2})}{w(x-i\frac{\gamma}{2})}\,u(x-i\gamma)\n
  &\quad+\sqrt{V(x+i\tfrac{M+2}{2}\gamma)V^*(x+i\tfrac{M}{2}\gamma)}\,
  \frac{w(x-i\frac{\gamma}{2})}{w(x+i\frac{\gamma}{2})}\,u(x+i\gamma)\n
  &\quad-\sqrt{V(x-i\tfrac{M}{2}\gamma)V^*(x-i\tfrac{M+2}{2}\gamma)}\,
  \frac{w(x+i\frac{\gamma}{2})}{w(x-i\frac{\gamma}{2})}
  \frac{w_1(x-i\gamma)}{w_1(x)}\,u(x)\n
  &\quad-\sqrt{V(x+i\tfrac{M+2}{2}\gamma)V^*(x+i\tfrac{M}{2}\gamma)}\,
  \frac{w(x-i\frac{\gamma}{2})}{w(x+i\frac{\gamma}{2})}
  \frac{w_1(x+i\gamma)}{w_1(x)}\,u(x)
  +\tilde{\mathcal{E}}_{d'}\,u(x).
  \label{id_a2}
\end{align}
By rewriting these \eqref{id_a2} and \eqref{id_i2} as
\begin{equation}
  \begin{pmatrix}
  \alpha_1a_1&\alpha_1^*a_1^*\\
  \alpha_2a_2\beta^*&\alpha_2^*a_2^*\beta
  \end{pmatrix}
  \begin{pmatrix}
  u(x-i\gamma)\\
  u(x+i\gamma)
  \end{pmatrix}
  =\begin{pmatrix}
  b_1u(x)\\
  b_2u(x)
 \end{pmatrix},
\end{equation}
where
\begin{align*}
  &\alpha_1=\sqrt{V(x-i\tfrac{M}{2}\gamma)V^*(x-i\tfrac{M+2}{2}\gamma)},\quad
  \alpha_2=\sqrt{V(x-i\tfrac{M+2}{2}\gamma)V^*(x-i\tfrac{M+4}{2}\gamma)},\\
  &a_1=\frac{w(x+i\frac{\gamma}{2})}{w(x-i\frac{\gamma}{2})}
  =\frac{1}{a_1^*},\quad
  a_2=\frac{w_3(x+i\frac{\gamma}{2})}{w_3(x-i\frac{\gamma}{2})}
  =\frac{1}{a_2^*},\quad
  c=\frac{w_1(x+i\gamma)}{w_1(x)},\\
  &\beta_1=\frac{g^P_{\mathcal{D}}(x+i\gamma)}{g^P_{\mathcal{D}}(x)},\quad
  \beta_2=\frac{g^P_{\mathcal{D}'''}(x+i\gamma)}{g^P_{\mathcal{D}'''}(x)},\quad
  \beta=\frac{\beta_2}{\beta_1},\\
  &b_1=\mathcal{E}_n-\tilde{\mathcal{E}}_{d'}
  +\alpha_1a_1c^*+\alpha_1^*a_1^*c,
  \ b_2=g^P_{\mathcal{D}}(x)\circ\widetilde{\mathcal{H}}_{\mathcal{D}'''}
  \circ g^P_{\mathcal{D}}(x)^{-1}-\tilde{\mathcal{E}}_{d''}
  +\alpha_1a_2^*c+\alpha_1^*a_2c^*,
\end{align*}
we obtain
\begin{equation}
  \begin{pmatrix}
  u(x-i\gamma)\\
  u(x+i\gamma)
  \end{pmatrix}
  =\frac{1}{\alpha_1\alpha_2^*a_1a_2^*\beta-\alpha_1^*\alpha_2a_1^*a_2\beta^*}
  \begin{pmatrix}
  \alpha_2^*a_2^*\beta&-\alpha_1^*a_1^*\\
  -\alpha_2a_2\beta^*&\alpha_1a_1
  \end{pmatrix}
  \begin{pmatrix}
  b_1u(x)\\
  b_2u(x)
 \end{pmatrix}.
 \label{ubyu}
\end{equation}
By using the explicit forms of $g^P_{\mathcal{D}}(x)$ and
$g^P_{\mathcal{D}'''}(x)$ ($d',d''$ : different types),
a lengthy but straightforward calculation shows the following:
\begin{equation}
  \alpha_1-\alpha_2^*\beta=0.
  \label{al1=al2*be}
\end{equation}

By writing down the identity \eqref{idmaru2} and multiplying by
\begin{equation*}
  \Bigl(\prod_{j=0}^{M+1}V\bigl(x+i(\tfrac{M+2}{2}-j)\gamma\bigr)
  V^*\bigl(x-i(\tfrac{M+2}{2}-j)\gamma\bigr)\Bigr)^{-\frac14}
  \sqrt{w_3(x-i\tfrac{\gamma}{2})w_3(x+i\tfrac{\gamma}{2})},
\end{equation*}
we obtain
\begin{align}
  u_3(x)
  &=-\frac{w_3(x+i\frac{\gamma}{2})}{w(x-i\frac{\gamma}{2})}u(x-i\gamma)
  -\frac{w_3(x-i\frac{\gamma}{2})}{w(x+i\frac{\gamma}{2})}u(x+i\gamma)\n
  &\quad+\frac{w_3(x-i\frac{\gamma}{2})}{w(x+i\frac{\gamma}{2})}cu(x)
  +\frac{w_3(x+i\frac{\gamma}{2})}{w(x-i\frac{\gamma}{2})}c^*u(x).
  \label{idmaru2u}
\end{align}
By substituting $u(x\mp i\gamma)$ in \eqref{ubyu} into \eqref{idmaru2u}
and using $u(x)=g^P_{\mathcal{D}}(x)\check{P}_{\mathcal{D},n}(x)$,
a short calculation gives
\begin{align*}
  &\quad(\alpha_1\alpha_2^*a_1a_2^*\beta-\alpha_1^*\alpha_2a_1^*a_2\beta^*)
  u_3(x)\\
  &=g^P_{\mathcal{D}}(x)
  \Bigl(\alpha_1^*\frac{w_3(x+i\frac{\gamma}{2})}{w(x+i\frac{\gamma}{2})}
  -\alpha_1\frac{w_3(x-i\frac{\gamma}{2})}{w(x-i\frac{\gamma}{2})}\Bigr)
  \bigl(\widetilde{\mathcal{H}}_{\mathcal{D}'''}
  +\mathcal{E}_n-\tilde{\mathcal{E}}_{d'}-\tilde{\mathcal{E}}_{d''}\bigl)
  \check{P}_{\mathcal{D},n}(x),
\end{align*}
where \eqref{al1=al2*be} is used.
Dividing this equation by $\alpha_1\alpha_1^*$ and using \eqref{al1=al2*be} and
$u_3(x)=g^P_{\mathcal{D}'''}(x)\check{P}_{\mathcal{D}''',n}(x)$, we obtain
\begin{align}
  &\quad\Bigl(\frac{\alpha_1}{\alpha_1^*}a_1a_2^*
  -\frac{\alpha_1^*}{\alpha_1}a_1^*a_2\Bigr)\check{P}_{\mathcal{D}''',n}(x)\n
  &=\frac{g^P_{\mathcal{D}}(x)}{g^P_{\mathcal{D}'''}(x)}
  \Bigl(\frac{1}{\alpha_1}
  \frac{w_3(x+i\frac{\gamma}{2})}{w(x+i\frac{\gamma}{2})}
  -\frac{1}{\alpha_1^*}
  \frac{w_3(x-i\frac{\gamma}{2})}{w(x-i\frac{\gamma}{2})}\Bigr)
  \bigl(\widetilde{\mathcal{H}}_{\mathcal{D}'''}
  +\mathcal{E}_n-\tilde{\mathcal{E}}_{d'}-\tilde{\mathcal{E}}_{d''}\bigl)
  \check{P}_{\mathcal{D},n}(x).
  \label{idmaru2u2}
\end{align}
By using the explicit forms of $g^P_{\mathcal{D}}(x)$,
$g^P_{\mathcal{D}'''}(x)$, $g_{\mathcal{D}}(x)$ and $g_{\mathcal{D}'''}(x)$,
a lengthy but straightforward calculation shows that \eqref{idmaru2u2} is
expressed as \eqref{idmaru2P}.
\hfill\fbox{}


\end{document}